\documentclass[11pt]{article}
\usepackage{amsmath}
\usepackage{amssymb}

\newtheorem{theorem}{Theorem}
\newtheorem{acknowledgement}[theorem]{Acknowledgement}

\input{tcilatex}
\begin{document}

\title{Worst Case in Voting and Bargaining}
\author{Anna Bogomolnaia\thanks{%
University of Glasgow, CNRS-CES, and Higher School of Economics, St
Petersburg; anna.bogomolnaia@glasgow.ac.uk}, Ron Holzman\thanks{%
Technion-Israel Institute of Technology. Work done during a visit at
Princeton University; holzman@technion.ac.il}, Herv\'{e} Moulin\thanks{%
University of Glasgow and Higher School of Economics, St Petersburg;
herve.moulin@glasgow.ac.uk}}
\maketitle

\begin{abstract}
The guarantee of an anonymous mechanism is the worst case welfare an agent
can secure against unanimously adversarial others. How high can such a
guarantee be, and what type of mechanism achieves it?

We address the worst case design question in the $n$-person probabilistic
voting/bargaining model with $p$ deterministic outcomes. If $n\geq p$ the
uniform lottery is the only maximal (unimprovable) guarantee; there are many
more if $p>n$, in particular the ones inspired by the random dictator
mechanism and by voting by veto.

If $n=2$ the maximal set $\mathcal{M}(n,p)$ is a simple polytope where each
vertex combines a round of vetoes with one of random dictatorship. For $%
p>n\geq 3$, writing $d=\lfloor \frac{p-1}{n}\rfloor $, we show that the dual
veto and random dictator guarantees, together with the uniform one, are the
building blocks of $2^{d}$ simplices of dimension $d$ in $\mathcal{M}(n,p)$.
Their vertices are guarantees easy to interpret and implement. The set $%
\mathcal{M}(n,p)$ may contain other guarantees as well; what we can say in
full generality is that it is a finite union of polytopes, all sharing the
uniform guarantee.

Keywords\textbf{: }\textit{worst case, guarantees, voting by veto, random
dictator}


\begin{acknowledgement}
We are grateful for the comments of seminar participants in the University
Paris 1, Lancaster, Oxford and Ashoka Universities. Moulin's research was
supported by a Blaise Pascal Chair of the Region Ile-de-France, 2020-21.
\end{acknowledgement}
\end{abstract}

\section{Guarantees and protocols}

Worst case analysis is a simple mechanism design question. Fix an arbitrary
collective decision problem by its feasible outcomes -- allocation of
resources, public decision making, etc.. --, the domain of individual
preferences and the number $n$ of relevant agents. We evaluate a mechanism
(game form) solving this problem by the \textit{guarantee} it offers to the
participants. This is the welfare level each one can secure in this game
form without any prior knowledge of how others will play their part: the
worst case assumption is that their moves are collectively adversarial (what
I know or believe about their preferences, what I expect about their
behaviour is irrelevant). My guarantee is the value of the two-person
zero-sum game pitting me against the rest of the world.

Given the decision problem, what guarantees can any mechanism offer, and
which mechanisms implement such guarantees? We are particularly interested
in the \textit{maximal} guarantees, those that cannot be improved: a higher
guarantee is a better default option if I am clueless about other
participants or unwilling to engage in risky strategic moves; it encourages
acceptance of and participation in the mechanism.

These questions were first addressed by the cake-cutting literature (\cite%
{St}, \cite{DS}, \cite{Kuh}). Two agents divide a cake over which their
utilities are additive and non atomic. In the Divide and Choose mechanism
(D\&C for short) they each can guarantee a share worth $1/2$ of the whole
cake: the Divider must cut the cake in two parts of equal worth, any other
move is at her own risk. This guarantee is not only maximal but also optimal
(higher than any other feasible guarantee): when the two agents have
identical preferences, their common guarantee cannot be worth more than $1/2$
of the cake.

Contrast D\&C with the simple Nash demand game: each agent claims a share of
the cake, demands are met if they are compatible, otherwise nobody gets any
cake. Against an adversarial player I will not get anything: this mechanism
offers no guarantee at all. The appeal of D\&C is that I cannot be tricked
to accept a share worth less than $1/2$.

Worst case analysis is related to the familiar implementation methodology in
mechanism design, but only loosely. We speak of a mechanism \textit{%
implementing} a certain guarantee -- mapping my preferences to a certain
welfare level -- but we do not postulate that each agent behaves under the
worst case assumption, nor do we ask what social choice function will then
be realised, as some papers reviewed in section 2 did. Instead the guarantee
of a game form is just one of its features, an important one for two reasons:

\begin{itemize}
\item an agent using a best reply to the other agents' strategies gets at
least her guaranteed welfare (because she has a safe strategy achieving that
level no matter what); so any Nash equilibrium of the game delivers at least
the guaranteed welfare to everyone,

\item if an agent plays the mechanism repeatedly with changing sets of
participants, the safe strategy is always available when she happens to be
clueless about the other agents' behaviour,
\end{itemize}

Many different mechanisms can implement the same guarantee, as the example
below makes clear.\footnote{%
Similarly we can implement the optimal \textquotedblleft one half of the
whole cake\textquotedblright\ guarantee by D\&C or by any one of Dubins and
Spanier' moving knife\textit{\ }procedures (\cite{DS})} We call the whole
class of game forms sharing a certain guarantee the \textit{protocol}
implementing it. The two concepts of guarantee and protocol and their
relation is the object of worst case analysis.

We initiate this approach in the probabilistic voting model, where the
protocols we identify can be interpreted as the guidelines for a partially
informal bargaining process. There are finitely many pure (deterministic)
outcomes and we must choose a convex compromise (probabilistic or otherwise)
between these.\ For tractability, we maintain a symmetric treatment of
agents (Anonymity) and of outcomes (Neutrality). We find that, depending on
the number $n$ of agents and $p$ of pure outcomes the set of maximal
guarantees and their protocols can be either very simple and dull (when $%
n\geq p$, see below) or dauntingly complex.

A good starting point is the simple case of deterministic voting over $p$
outcomes with ordinal preferences. An anonymous and neutral guarantee is a
rank $k$ from $1$ to $p$ (where rank $1$ is the worst): it is feasible if
for any preference profile there is at least one outcome ranked $k$ or above
by each voter. Suppose first $n\geq p$: at a profile where each outcome is
the worst for some agent, the rank $k$ must be $1$, so the guarantee idea
has no bite. Now if $n<p$ we can give to each voter the right to veto up to $%
d=\lfloor \frac{p-1}{n}\rfloor $ outcomes: this is feasible because $nd<p$,
so the rank $k=d+1$ is a feasible guarantee, clearly the best possible one.
Worst case analysis' simple advice to a committee smaller than the number of
outcomes it chooses from, is to distribute $d$ veto rights to its members,
not at all what a standard voting rule \`{a} la Condorcet or Borda does.
However the corresponding protocol contains, inter alia, the following
mechanisms: ask everyone to pick independently $d$ outcomes to veto, then
use any voting rule to pick among the remaining free outcomes, often more
than $p-nd$ of them.

We allow compromises between the pure outcomes, interpreted as lotteries,
time shares, or the division of a budget. Distributing veto tokens is a
natural way to achieve a high guarantee, but there are others. The familiar
random dictator\footnote{%
Each agent has an equal chance to choose the final outcome.} mechanism (\cite%
{Gib}) with two voters implements the guarantee putting a $1/2$ probability
on both my first and worst ranks. And the uniform lottery over all ranks is
yet another guarantee implemented by any mechanism where each participant
has the right, at some stage of the game that could depend upon the agent,
the play of the game etc.., to force the decision by flipping a fair coin
between all outcomes.\smallskip

\paragraph{\textit{Example: three agents, six outcomes}}

\noindent The uniform guarantee $UNI(6)$ is the lottery $\lambda ^{uni}=(%
\frac{1}{6},\frac{1}{6},\frac{1}{6},\frac{1}{6},\frac{1}{6},\frac{1}{6})$,
where each rank is equally probable.

Distributing one veto token to each agent implements $\lambda
^{1}=(0,1,0,0,0,0)$ (recall the first coordinate is the worst rank), as in
the deterministic case. But $\lambda ^{1}$ is not maximal: it is improved by
making the protocol a bit more precise. After the veto tokens have been
used, we can pick one of the remaining outcomes uniformly, or give the
option to force this random choice to each agent. Then the rank distribution
cannot be worse for anyone than $\lambda ^{vt}=(0,\frac{1}{3},\frac{1}{3},%
\frac{1}{3},0,0)$, because my worst case after the vetoing phase is that the
two other agents killed my two best outcomes. And $\lambda ^{vt}$
stochastically dominates $\lambda ^{1}$. We will use the notation $VT(3,6)$
instead of $\lambda ^{vt}$ when it is important to specify $n$ and $p$.

The random dictator mechanism between our three agents delivers the
guarantee $\lambda ^{2}=(\frac{2}{3},0,0,0,0,\frac{1}{3})$: my worst case is
that the two other agents pick my worst outcome. Again $\lambda ^{2}$ is not
maximal, and improved by the following protocol: agents report (one of)
their top outcome(s); if they all agree on $a$ we choose $a$; if they each
choose a different outcome, we pick one of them with uniform probability;
but if the choices are $a,a,b$ we randomize uniformly between $a,b$ \textit{%
and an arbitrary third outcome} $c$. This implements the correct random
dictator guarantee $RD(3,6):$ $\lambda ^{rd}=(\frac{1}{3},\frac{1}{3},0,0,0,%
\frac{1}{3})$, that stochastically dominates $\lambda ^{2}$.

It is easy to check directly that $UNI(6)$, $VT(3,6)$ and $RD(3,6)$ are all
maximal. For instance this follows for $\lambda ^{rd}$ and $\lambda ^{vt}$
by inspecting respectively the left or right profile of strict ordinal
preferences%
\begin{equation*}
\begin{array}{ccccccc}
\prec _{1} & a & b & x & y & z & c \\ 
\prec _{2} & b & c & y & z & x & a \\ 
\prec _{3} & c & a & z & x & y & b%
\end{array}%
\text{ \ }%
\begin{array}{ccccccc}
\prec _{1} & a & x & y & z & b & c \\ 
\prec _{2} & b & y & z & x & c & a \\ 
\prec _{3} & c & z & x & y & a & b%
\end{array}%
\end{equation*}%
(where agent $1$'s worst is $a$ and best is $c$). At the left profile, to
give a $\frac{1}{3}$ chance of their best outcome to all agents a protocol
implementing $\lambda ^{rd}$ must pick $a,b$ or $c$, each with probability $%
\frac{1}{3}$: then each agent experiences exactly the distribution $\lambda
^{rd}$ over her ranked outcomes, and no other lottery $\lambda $
stochastically dominating $\lambda ^{rd}$ is a feasible guarantee at this
profile. Similarly at the right profile, implementing $\lambda ^{vt}$
implies zero probability on $a,b,c$, and at most (hence exactly) $\frac{1}{3}
$ on each of $x,y$ and $z$. The symmetry of these two arguments is not a
coincidence: a critical duality relation connects $\lambda ^{vt}$ and $%
\lambda ^{rd}$ (section 4).

What other guarantees are maximal for $n=3,p=6$? Convex combinations
preserve feasibility but not maximality: for instance an equal chance of the
protocols implementing $VT(3,6)$ and $RD(3,6)$ delivers the feasible
guarantee $\frac{1}{2}\lambda ^{rd}+\frac{1}{2}\lambda ^{vt}=(\frac{1}{6},%
\frac{1}{3},\frac{1}{6},\frac{1}{6},0,\frac{1}{6})$ which is dominated by $%
\lambda ^{uni}$. But lotteries between $UNI(6)$ and $VT(3,6)$, or between $%
UNI(6)$ and $RD(3,6)$ are in fact maximal. Moreover for this choice of $n$
and $p$, the maximal guarantees cover exactly the two intervals $[\lambda
^{uni},\lambda ^{vt}]$ and $[\lambda ^{uni},\lambda ^{rd}]$ (Theorem 1 in
section 5).

The choice facing the worst case designer in this example is sharp, and its
resolution is context dependent: the veto guarantee is a good fit when
bargaining is about choosing an expensive infrastructure project, or a
person to hold a position for life; the random dictator approach makes sense
if we are dividing time between different activities, or choosing a pair of
roman consuls; the uniform guarantee stands out if we value a disagreement
outcome revealing no information about individual preferences.

Critical to their practical application, the protocols implementing $UNI,VT$
and $RD$ above rely on ordinal preferences only, as do the agents' safe
action when they report which outcome(s) they veto, or which ones they
prefer among those still in play.

\paragraph{The punchline}

Our results cast a new light on two familiar collective decison mechanisms,
random dictator and voting by veto. Together and in combination with the
uniform guarantee, they generate all maximal guarantees if $n=2$, and
essentially all of them again if $3\leq n<p\leq 2n$. In the general case
they can be sequentially combined to produce a very large subset of maximal
guarantees.

\subsection{Contents of the paper}

After a review of the literature in section 2 we define in section 3 the
concept of guarantee in three related models. In the first one, agents have
ordinal preferences over the pure outcomes, and incomplete preferences over
lotteries by stochastic dominance. In the second they have von Neuman
Morgenstern (vNM) utilities over lotteries. In the third they have
quasi-linear utilities over outcomes and money, and lotteries are replaced
by cash compensations. A guarantee is a convex combination of the ranks $1$
to $p$ where rank $1$ is the worst. It is feasible if at each profile of
preferences, there is a lottery over pure outcomes, or a pure outcome and a
balanced set of cash compensations in the quasi-linear model, that everyone
weakly prefers to her guaranteed utility.

Lemma 1 shows that the three definitions are equivalent and that feasible
guarantees cover a canonical polytope $\mathcal{G}(n,p)$ in the simplex with 
$p$ ranked vertices. Its Corollary gives a compact though abstract
characterisation of $\mathcal{G}(n,p)$.

Section 4 focuses on the subset $\mathcal{M}(n,p)$ of maximal guarantees,
starting with a complete characterisation in two easy cases (section 4.1):

If $n\geq p$ the unique maximal guarantee is $UNI(p)$, dominating every
other feasible guarantee (Proposition 1), so the worst case viewpoint tells
us to allow each agent to force this canonical anonymous and neutral
disagreement outcome. In every other case there are many more options.

If $n=2<p$ a guarantee $\lambda $ is maximal if and only if it is symmetric
with respect to the middle rank (Proposition 2). For instance $\mathcal{M}%
(2,6)$ is the convex hull of $\lambda ^{rd}=(\frac{1}{2},0,0,0,0,\frac{1}{2}%
) $ ($RD(2,6)$), $\lambda ^{mix}=(0,\frac{1}{2},0,0,\frac{1}{2},0)$, and $%
\lambda ^{vt}=(0,0,\frac{1}{2},\frac{1}{2},0,0)$ (two veto tokens per
person). Here is the protocol for $\lambda ^{mix}$: the agents veto one
outcome each, then choose randomly a dictator (equivalently, we randomly
give one veto token to one agent and four tokens to the other). Note that $%
UNI(p)$ is the center of the polytope $\mathcal{M}(2,p)$.

When $3\leq n<p$, the structure of $\mathcal{M}(n,p)$ is much more
complicated.\ Section 4.2 describes a critical duality property inside $%
\mathcal{M}(n,p)$, relating $VT(n,p)$ and $RD(n,p)$, while $UNI(p)$ is
self-dual: Proposition 3. We define in section 4.3 the large set $\mathcal{C}%
(n,p)$ of \textit{canonical guarantees}: for three or more agents they are a
rich family of vertices of $\mathcal{M}(n,p)$. Their protocols\textit{\ }%
combine up to $d$ successive rounds (recall $d=\lfloor \frac{p-1}{n}\rfloor $%
) of either veto (one token each) or a (partial) random dictator.

Our first main result Theorem 1 in section 5.1, gives a fairly complete
picture of all maximal guarantees with three or more agents and at most
twice as many pure outcomes ($p\leq 2n\Longleftrightarrow d=1$). As long as $%
p\neq 2n-1$ and $(n,p)\neq (4,8),(5,10)$, they cover exactly the two
intervals $[UNI(p),VT(n,p)]$ and $[UNI(p),RD(n,p)]$, as in the numerical
example above. There are additional maximal\ guarantees when $p=2n-1$ or $%
(n,p)=(4,8),(5,10)$, some of them described after the Theorem (Proposition
4).

In section 5.2 we turn to the general case $3\leq n<p$ with no restrictions
on $d$. The set $\mathcal{M}(n,p)$ is a union of polytopes (faces of $%
\mathcal{G(}n,p)$), of which $UNI(p)$ is always a vertex. Theorem 2 uses the
canonical guarantees in $\mathcal{C}(n,p)$ to construct $2^{d}$ simplices of
dimension $d$, one for each sequence $\Gamma $ of length $d$ in $\{VT,RD\}$:
the vertices of such a simplex are lotteries in $\mathcal{C}(n,p)$ obtained
from the $d$ initial subsequences of $\Gamma $, plus $UNI(p)$. For instance
the sequence $\Gamma =(VT,RD)$ gives the triangle in $\mathcal{C}(3,7)$ with
vertices $UNI(p)$, $VT(3,7)$, and $\lambda =(0,\frac{1}{3},\frac{1}{3},0,%
\frac{1}{3},0,0)$ denoted $VT\otimes RD$; the latter is implemented by a
first round of one veto each, followed by $RD(3,q)$ over the remaining $q$
outcomes (four or more). This construction does not cover the entire set $%
\mathcal{M}(n,p)$ but delivers a large subset built from simple combinations
of veto and random dictator steps.

Section 6 gathers some open questions and potential research directions.
Several proofs are gathered in the Appendix, section 7.

\section{Related literature}

The optimal design of a mechanism under the risk averse assumption that
other agents are adversarial is discussed by the early literature on
implementation in several slightly different formulations: implementation in
maximin (\cite{Th1}, \cite{DHM}), prudent (\cite{Mou2}) or protective
strategies (\cite{BD}). As explained in section 1 our protocols do not
define complete game forms, and our guarantees are compatible with a wide
range of strategic behaviours.

Steinhaus' seminal papers (\cite{St}, see also \cite{DS}, \cite{Kuh})
invented the worst case approach for cutting a cake fairly among any number
of agents. His simple protocol generalises Divide and Choose and guarantees
to each agent a \textit{fair share}: one that is worth at least $\frac{1}{n}$
of the whole cake. The main focus of the subsequent literature is envy free
divisions: how to achieve one by simple cuts and queries (\cite{BT1}, \cite%
{RW}, \cite{AMK}) and proving its existence under preferences more general
than additive utilities (\cite{Str}, \cite{Woo}). An exception is the recent
paper (\cite{BM1}) returning to the worst case approach under very general
preferences and identifying the MinMaxShare (my best share in the worst
partition of the cake I can be offered) as a feasible guarantee, though not
a maximal one.

The last decade saw an explosion of research to define and compute a fair
allocation of indivisible items, proposing in particular a new definition of
the fair share as the MaxMinShare (\cite{Bud}): my worst share in the best
partition of the objects I can propose. This guarantee may not be feasible (%
\cite{PW}) but this happens very rarely (\cite{KPW}); the real concern is
that the protocols approximating this guarantee are all but simple.

Other early instances of the worst case approach are in production economies
(\cite{Mou3}, \cite{Mou4}) and in the minimal cost spanning tree problem (%
\cite{HMO}).

The random dictator mechanism is a staple of probabilistic social choice (%
\cite{Gib}, \cite{Sen}). In axiomatic bargaining it inspires the\ Raiffa
solution (\cite{Rai}) and the mid-point domination axiom (\cite{Sob}, \cite%
{Th}) satisfied by both the Nash and Kalai-Smorodinsky solutions.

Voting by veto is another early idea introduced by Mueller (\cite{Mue}) to
incentivise agents toward compromising offers: each agent makes one offer,
which together with the status quo outcome makes $p=n+1$ outcomes, after
which they take turns to veto one outcome each (in our model the natural
status quo is the uniform lottery over outcomes). This procedure is
generalised in (\cite{Mou2}). The area monotonic bargaining solution (\cite%
{ABig}, \cite{Anb}) is a direct application of voting by veto between two
parties, similar to distributing $\lfloor \frac{p-1}{2}\rfloor $ veto tokens
to each agent in our model.

A handful of recent papers discuss variants of voting by veto in the classic
implementation context: (\cite{CEK}), (\cite{BC}),(\cite{LNS}). All three
papers implement maximal guarantees\textbf{. }Closer to home section 4 in (%
\cite{KiN}) explains the strategic properties of a veto mechanism
implementing arbitrary compositions of our guarantee $VT(n,p)$.

We mention finally the small literature on bargaining with cash
compensations and quasi-linear utilities (\cite{Mou1}, \cite{Chu}) where
only the uniform guarantee is discussed, while our results unveil many more
possibilities.

\section{Feasible guarantees}

Anonymity and Neutrality (symmetric treatment of agents and outcomes,
respectively) are hard wired in the model so a guarantee is well defined
once we fix the number $n$ of agents and $p$ of deterministic outcomes. It
is an element $\lambda $ of $\Delta (p)$, the simplex of lotteries over the
ranks in $[p]=\{1,\cdots ,p\}$. Here $\lambda _{1}$ is the probability of
the worst rank and $\lambda _{p}$ that of the best rank. We give three
equivalent definitions of the same concept of feasibility, after which when
we speak of a guaranteee we always mean that it is feasible.

Notation. For lotteries $\lambda \in \Delta (p)$, and only for those, we
write $[\lambda ]_{k_{1}}^{k_{2}}$ instead of the sum $\sum_{k_{1}}^{k_{2}}%
\lambda _{t}$. The symmetric of $\lambda $ w.r.t. the middle rank is $%
\widetilde{\lambda }$: $\widetilde{\lambda }_{k}=\lambda _{p+1-k}$ for all $%
k\in \lbrack p]$.

The stochastic dominance relation (dominance for short) in $\Delta (p)$
plays a central role throughout.\textbf{\ }We write $\lambda \vdash \mu $
and say that $\lambda $ dominates $\mu $ if the following three equivalent
properties hold%
\begin{equation*}
\forall k\in \lbrack p]:[{\large \lambda }]_{1}^{k}\leq \lbrack {\large \mu }%
]_{1}^{k}
\end{equation*}%
\begin{equation*}
\forall k\in \lbrack p]:[{\large \lambda }]_{k}^{p}\geq \lbrack {\large \mu }%
]_{k}^{p}
\end{equation*}
\begin{equation*}
\forall z\in 
\mathbb{R}
^{p}:\{z_{1}\leq z_{2}\leq \cdots \leq z_{p}\}\Longrightarrow \lambda \cdot
z\geq \mu \cdot z
\end{equation*}

The set of deterministic outcomes is $A$, with generic element $a$, and $%
\Delta (A)$, with generic element {\large $\ell $}, is that of lotteries
over $A$. We keep in mind the alternative interpretations of $\Delta (A)$ as
time sharing or division of a budget between the \textquotedblleft
pure\textquotedblright\ outcomes in $A$.

The set of agents is $\lbrack n]$, with generic element $i$. An agent $i$'s
ordinal preference over $A$ (a complete, reflexive and transitive relation)
is written $\succsim _{i}$. A $k$-tail of the preference $\succsim _{i}$ is
a subset $T$ of $A$ with cardinality $k$ such that $b\succsim _{i}a$
whenever $a\in T,b\in A\diagdown T$. \ Indifferences in $\succsim _{i}$ may
produce several $k$-tails.

Given $\succsim _{i}$ and ${\large \ell }\in \Delta (A)$ the rank-ordered
rearrangement of ${\large \ell }$ is the following lottery ${\large \ell }%
^{\ast i}$ in $\Delta (p)$%
\begin{equation}
\forall k\in \lbrack p]:[{\large \ell }^{\ast i}]_{1}^{k}=\min \{\sum_{a\in
T}{\large \ell }_{a}|T\text{ is a }k\text{-tail of}\succsim _{i}\}
\label{11}
\end{equation}

\textbf{Definition 1 (ordinal preferences): }\textit{Given }$n$ and $p$, 
\textit{the lottery }$\lambda \in \Delta (p)$\textit{\ is a guarantee at }$%
n,p$\textit{\ if}\textbf{\ }\textit{for any }$n$-\textit{profile of
preferences }$\pi =(\succsim _{i})_{i=1}^{n}$\textit{\ on }$A$\textit{\
there exists a lottery }{\large $\ell $}$\in \Delta (A)$\textit{\ s.t. }%
{\large $\ell $}$^{\ast i}\vdash \lambda $\textit{\ for all }$i\in \lbrack
n] $\textit{. Then we say that the lottery }{\large $\ell $}\textit{\
implements }$\lambda $\textit{\ at profile }$\pi $\textit{.\smallskip }

An agent $i$'s vNM utility over $A$ is a vector $u_{i}$ in $%
\mathbb{R}
^{A}$ and $u_{i}\cdot {\large \ell }=\sum_{a\in A}u_{ia}{\large \ell }_{a}$
is her utility at lottery ${\large \ell }$. We write $u_{i}^{\ast }\in 
\mathbb{R}
^{p}$ the rank-ordered rearrangement (aka order statistics) of $u_{i}$:%
\begin{equation*}
\forall k\in \lbrack p]:\sum_{t=1}^{k}{\large u}_{it}^{\ast}=\min
\{\sum_{a\in T}{\large u}_{ia}|T\subseteq A\text{, }|T|=k\}
\end{equation*}

\textbf{Definition 2 (vNM utilities):}\textit{Given }$n$ and $p$, \textit{%
the lottery }$\lambda \in \Delta (p)$\textit{\ is a guarantee at }$n,p$%
\textit{\ if}\textbf{\ }\textit{for any }$n$-\textit{profile of utilities }$%
(u_{i})_{i=1}^{n}$ \textit{on }$A$\textit{\ there exists a lottery }${\large %
\ell }\in \Delta (A)$\textit{\ s.t. }${\large \ell }\cdot u_{i}\geq \lambda
\cdot u_{i}^{\ast }$\textit{\ for all }$i\in \lbrack n]$\textit{.\smallskip }

The ordinal definition is agnostic w.r.t. the risk attitude of the agents.
The cardinal one specifies it completely.

In the third model the agents have quasi-linear utilities over the pure
outcomes in $A$: in lieu of randomisation (or any convex combinations)
compromises are achieved by cash compensations between the agents. Agent $i$%
's utility is still any $u_{i}\in 
\mathbb{R}
^{A}$ and an outcome is a pair $(a,t)$ where $a\in A$ and $%
t=(t_{i})_{i=1}^{n}$ is a balanced set of transfers between agents, $%
\sum_{1}^{n}t_{i}=0$; the corresponding utilities are $u_{ia}+t_{i}$%
.\smallskip

\textbf{Definition 3 (quasi-linear utilities): }\textit{Given }$n$ and $p$, 
\textit{the convex combination }$\lambda \in \Delta (p)$\textit{\ is a
guarantee at }$n,p$\textit{\ if}\textbf{\ }\textit{for any }$n$-\textit{%
profile }$(u_{i})_{i=1}^{n}$\textit{\ of utilities on }$A$, \textit{there
exists an outcome $a \in A$ and a balanced set of transfers }$%
(t_{i})_{i=1}^{n}$\textit{\ such that} $u_{ia}+t_{i}\geq \lambda \cdot
u_{i}^{\ast }$\textit{\ for all }$i\in \lbrack n]$\textit{.\smallskip }

\textbf{Lemma 1 }\textit{These three definitions are equivalent.}

\noindent \textit{We write }$\mathcal{G}(n,p)$ \textit{for the set of all
guarantees at }$n,p$\textit{: it is a polytope in }$\Delta (p)$\textit{%
.\smallskip }

\textbf{Proof}

\noindent \textit{\ Definition 1 }$\Longrightarrow $\textit{\ Definition 2}

The identity ${\large \ell }\cdot u_{i}={\large \ell }^{\ast i}\cdot
u_{i}^{\ast }$ for all ${\large \ell }\in \Delta (A),u_{i}\in 
\mathbb{R}
^{A}$ is easily checked. Now assume $\lambda $ meets Definition 1 and fix an
arbitrary profile $(u_{i})_{i=1}^{n}$ of vNM utilities, with associated
ordinal preferences $(\succsim _{i})_{i=1}^{n}$. If {\large $\ell $}\textit{%
\ }implements $\lambda $ at $(\succsim_i)_{i=1}^n$ the relation ${\large %
\ell }^{\ast i}\vdash \lambda $ and the identity give ${\large \ell }\cdot
u_{i}\geq \lambda \cdot u_{i}^{\ast }$ as desired.\smallskip

\noindent \textit{\ Definition 2 }$\Longrightarrow $\textit{\ Definition 3}

Definition 3 says that $\lambda $\textit{\ }is a guarantee if and only if
for all $(u_{i})_{i=1}^{n}\in (%
\mathbb{R}
^{A})^{n}$ we have:%
\begin{equation}
\sum_{i=1}^{n}\lambda \cdot u_{i}^{\ast }\leq \max_{a\in
A}\sum_{i=1}^{n}u_{ia}  \label{1}
\end{equation}%
Fix $(u_{i})_{i=1}^{n}$ and choose {\large $\ell $}\textit{\ }implementing $%
\lambda $ as in Definition 2: the inequalities ${\large \ell }\cdot
u_{i}\geq \lambda \cdot u_{i}^{\ast }$ and 
\begin{equation*}
\sum_{i=1}^{n}{\large \ell }\cdot u_{i}\leq \max_{a\in A}\sum_{i=1}^{n}u_{ia}
\end{equation*}%
together imply (\ref{1}).\smallskip

\noindent\ \textit{Definition 3 }$\Longrightarrow $\textit{\ Definition 1. }%
Fix $\lambda \ $as in Definition 3 and a preference profile $(\succsim
_{i})_{i=1}^{n}$. Call $S_{i}$ the set of utilities $v_{i}\in 
\mathbb{R}
^{A}$ representing $\succsim _{i}$ ($a\succsim _{i}b\Longleftrightarrow
v_{ia}\geq v_{ib}$ for all $a,b$) and such that $v_{ia}\in \lbrack 0,1]$ for
all $a$. By property (\ref{1}) for any profile $(v_{i})_{i=1}^{n}\in {\large %
\Pi }_{i=1}^{n}S_{i}$ there exists $a \in A$ such that $\sum_{i=1}^{n}v_{ia}%
\geq \sum_{i=1}^{n}\lambda \cdot v_{i}^{\ast }$, which implies%
\begin{equation*}
\min_{(v_{i})_{i=1}^{n}\in {\large \Pi }_{i=1}^{n}S_{i}}\max_{{\large a \in A%
}}\sum_{i=1}^{n}(v_{ia}-\lambda \cdot v_{i}^{\ast })\geq 0
\end{equation*}%
The summation is a linear function of the variable $(v_{i})_{i=1}^{n}$
varying in a convex compact, and of $a$. By the minimax theorem there exists 
${\large \ell }\in \Delta (A)$ such that $\sum_{i=1}^{n}{\large \ell }\cdot
v_{i}\geq \sum_{i=1}^{n}\lambda \cdot v_{i}^{\ast }$ for all $%
(v_{i})_{i=1}^{n}\in {\large \Pi }_{i=1}^{n}S_{i}$. Taking $v_{i}=0$ for all 
$i\geq 2$ gives ${\large \ell }\cdot v_{1}\geq \lambda \cdot v_{1}^{\ast }$
for all $v_{1}\in S_{1}$. Equivalently ${\large \ell }^{\ast 1}\cdot
v_{1}^{\ast }\geq \lambda \cdot v_{1}^{\ast }$ for any weakly increasing
sequence $v_{1}^{\ast }$ in $[0,1]^{p}$: the desired property ${\large \ell }%
^{\ast 1}\vdash \lambda $ follows, and the argument is the same for each $%
i\geq 2$. \smallskip

\noindent\ \textit{$\mathcal{G}(n,p)$ is a polytope.} Definition 1 implies
that the set $\mathcal{G}(n,p)$ is a polytope in $\Delta (p)$ (the convex
hull of a finite set, and the intersection of finitely many half-spaces) for
any $n,p$. Indeed feasibility of $\lambda $ at some fixed ordinal profile $%
\pi $ means that a system of linear inequalities in $\ell $ of the form $%
M\ell \geq \theta $ has a solution $\ell $, where the matrix $M$ is
independent of $\lambda $ and $\theta $ depends affinely on $\lambda $: by
the Farkas Lemma this is equivalent to finitely many linear inequalities on $%
\lambda $, and there are only finitely many ordinal profiles. $\blacksquare $%
\smallskip

\textbf{Corollary to Lemma 1 }\textit{The lottery }$\lambda \in \Delta (p)$%
\textit{\ is in }$\mathcal{G}(n,p)$\textit{\ if and only if for any }$n$-%
\textit{profile }$(u_{i})_{i=1}^{n}$ in $(%
\mathbb{R}
^{A})^{n}$\textit{\ we have}%
\begin{equation}
\sum_{i=1}^{n}u_{i}=0\Longrightarrow \sum_{i=1}^{n}\lambda \cdot u_{i}^{\ast
}\leq 0  \label{2}
\end{equation}
\textquotedblleft Only if\textquotedblright\ holds because (\ref{2}) is a
special case of the characteristic property (\ref{1}). For \textquotedblleft
if\textquotedblright\ we pick an arbitrary profile $(u_{i})_{i=1}^{n}$ and
set $z=\max_{a\in A}\sum_{i=1}^{n}u_{ia}$. Writing ${\boldsymbol{1}}$ the
vector with all coordinates equal to $1$, we pick a profile $%
(v_{i})_{i=1}^{n}$ such that $u_{i}\leq v_{i}$ for all $i$ and $%
\sum_{i=1}^{n}v_{ia}=z$ for all $a$, then we can apply (\ref{2}) to $%
(w_{i})_{i=1}^{n}$: $w_{i}=v_{i}-\frac{z}{n}{\boldsymbol{1}}$. This gives $%
\sum_{i=1}^{n}\lambda \cdot v_{i}^{\ast }\leq z $ and the claim.$\smallskip $

If $n=2$ property (\ref{2}) is easy to interpret, after checking the
following identity (recall $\widetilde{\lambda }$ is the symmetric of $%
\lambda $ w.r.t. the middle rank):%
\begin{equation}
\forall u\in 
\mathbb{R}
^{A}:\lambda \cdot (-u)^{\ast }=-\widetilde{\lambda }\cdot u^{\ast }
\label{15}
\end{equation}%
Property (\ref{2}) means $\lambda \cdot u^{\ast }\leq \widetilde{\lambda }%
\cdot u^{\ast }$ for all $u$, equivalently $\widetilde{\lambda }\vdash
\lambda $. Therefore%
\begin{equation}
\lambda \in \mathcal{G}(2,p)\Longleftrightarrow \lbrack \lambda
]_{1}^{k}\geq \lbrack \lambda ]_{p+1-k}^{p}\text{ for all }k=1,\cdots
,\lfloor \frac{p}{2}\rfloor  \label{12}
\end{equation}

But for $n\geq 3$ it is much harder to discover a set of such inequalities
representing $\mathcal{G}(n,p)$, or the set of its extreme points. \smallskip

\textit{Remark.} \textit{We downplay the quasi-linear interpretation because
the corresponding protocols, though quite straightforward to construct, are
less palatable. They require to report the willingness to pay for different
outcomes, and thus essentially rely on the entire cardinal utility profile. }

\section{Maximal guarantees}

From the welfare point of view, the guarantees of interest are those that
cannot be improved, the maximal ones.\smallskip

\textbf{Definition 4} \textit{The guarantee }$\lambda \in \mathcal{G}(n,p)$ 
\textit{is maximal if}%
\begin{equation*}
\forall \mu \in \mathcal{G}(n,p):\mu \vdash \lambda \Longrightarrow \mu
=\lambda
\end{equation*}%
\textit{The set of maximal guarantees is} $\mathcal{M}(n,p)\subset \mathcal{G%
}(n,p)$.

\subsection{Two easy cases: $n\geq p$ and $n=2$}

\textbf{Proposition 1 }\textit{The uniform guarantee }$UNI(p)$,\textit{\ }$%
\lambda _{k}^{uni}=\frac{1}{p}$\textit{\ for all} $k\in \lbrack p]$, has the
following properties:

\noindent $i)$\textit{\ It} \textit{is maximal for all} $n,p$.

\noindent $ii)$ \textit{If }$n\geq p$\textit{\ it dominates every other
feasible guarantee:} $\mathcal{M}(n,p)=\{\lambda ^{uni}\}$.

\noindent $iii)$ \textit{If} $n\geq 3$\textit{\ it is a vertex of} $\mathcal{%
G}(n,p)$\textit{, hence of} $\mathcal{M}(n,p)$ \textit{too}.\smallskip

\textbf{Proof.}

\textit{Statement }$i)$.The equality $\lambda ^{uni}\cdot u_{i}^{\ast
}=\lambda ^{uni}\cdot u_{i}$ for all $u_{i}$ implies for any profile $%
(u_{i})_{i=1}^{n}$%
\begin{equation}
\sum_{i=1}^{n}u_{i}=0\Longrightarrow \lambda ^{uni}\cdot {\large (}%
\sum_{i=1}^{n}u_{i}^{\ast }{\large )}=0  \label{16}
\end{equation}

Suppose some $\mu \in \mathcal{G}(n,p)$ dominates $\lambda ^{uni}$ and
consider a profile of the form $(u_{1},-u_{1},0,\cdots ,0)$ where $u_{1}$ is
arbitrary. Summing up the inequalities $\mu \cdot u_{1}^{\ast }\geq \lambda
^{uni}\cdot u_{1}^{\ast }$ , $\mu \cdot (-u_{1})^{\ast }\geq \lambda
^{uni}\cdot (-u_{1})^{\ast }$ gives $\mu \cdot u_{1}^{\ast }+\mu \cdot
(-u_{1})^{\ast }\geq 0$. Because $\mu $ meets property (\ref{2}) both
inequalities are equalities, and we conclude $\mu =\lambda^{uni} $.\smallskip

\textit{Statement }$ii)$. Assume $n\geq p$ and pick an arbitrary guarantee $%
\lambda $ in $\mathcal{G}(n,p)$, and a cyclical permutation $\sigma $ of $A$%
: the latter maps utility $u$ to $u^{\sigma }$: $u_{a}^{\sigma }=u_{\sigma
(a)}$. We pick any $u$ and consider the profile 
\begin{equation*}
(u,u^{\sigma },u^{\sigma ^{2}},\cdots ,u^{\sigma ^{p-1}},0,\cdots ,0)
\end{equation*}
with $n-p$ null utilities. Clearly $\sum_{k=0}^{p-1}u^{\sigma ^{k}}=\gamma 
{\large 1}$ for $\gamma =\sum_{a\in A}u_{a}$, so we can apply property (\ref%
{2}) to the profile $(u-\frac{\gamma }{p}{\large 1},u^{\sigma }-\frac{\gamma 
}{p}{\large 1},\cdots ,u^{\sigma ^{p-1}}-\frac{\gamma }{p}{\large 1}%
,0,\cdots ,0)$. Together with $(u^{\sigma ^{k}})^{\ast }=u^{\ast }$ for all $%
k$, this gives%
\begin{equation*}
0\geq \sum_{k=0}^{p-1}{\large (}\lambda \cdot (u^{\sigma ^{k}})^{\ast }-%
\frac{\gamma }{p}{\large )=p(}\lambda \cdot u^{\ast })-\gamma
\Longrightarrow \lambda \cdot u^{\ast }\leq \frac{\gamma }{p}=\lambda
^{uni}\cdot u^{\ast }
\end{equation*}%
as desired.\smallskip

\textit{Statement }$iii)$. Suppose $\lambda ^{uni}$ is a convex combination
of two distinct $\lambda ^{1},\lambda ^{2}$ in $\mathcal{G}(n,p)$. For any
profile s. t. $\sum_{i=1}^{n}u_{i}=0$, property (\ref{2}) implies $%
\sum_{i=1}^{n}\lambda ^{s}\cdot u_{i}^{\ast }\leq 0$ for $s=1,2$. But by (%
\ref{16}) the relevant convex combination of these inequalities is an
equality, therefore they both are equalities.

For $n\geq 3$, a lottery $\lambda $ meeting (\ref{16}) must be $\lambda
^{uni}$. Indeed, define $\delta _{k}=[\lambda ]_{p+1-k}^{p}$ for $k\in
\lbrack p]$ and $\delta _{0}=0$, then pick any three nonnegative integers $%
k,l,m$ summing to $p$. Consider a profile of $0,1$ utilities where $%
u_{1};u_{2};u_{3}$ are equal to $1$ precisely on three sets of respective
sizes $k,l,m$ partitioning $A$, while other utilities, if any, are
identically zero. Applying (\ref{16}) to this profile yields $\delta
_{k}+\delta _{l}+\delta _{m}=1$. It is easy to check that this simple
integer version of the Cauchy equation implies $\delta _{k}=\frac{k}{p}$ for
all $k$. $\blacksquare \smallskip $

\textbf{Proposition 2 }\textit{Maximal guarantees for }$n=2$

\noindent \textit{If }$n=2<p$\textit{\ the lottery }$\lambda \in \Delta (p)$%
\textit{\ is a maximal guarantee if and only if it is symmetric:}%
\begin{equation}
\lambda _{k}=\lambda _{p+1-k}\text{ for }1\leq k\leq \lfloor \frac{p}{2}%
\rfloor  \label{4}
\end{equation}%
\textit{The extreme points of the polytope }$\mathcal{M}(2,p)$ \textit{are
the following guarantees }$\lambda ^{t}$\textit{:}%
\begin{equation*}
\lambda _{t}^{t}=\lambda _{p+1-t}^{t}=\frac{1}{2}\text{ for }t=1,\cdots
,\lfloor \frac{p}{2}\rfloor \text{ ; and }\lambda _{\frac{p+1}{2}}^{\frac{p+1%
}{2}}=1\text{ if }p\text{ is odd}
\end{equation*}

We see that for $n=2$ the uniform guarantee $UNI(p)$ is the center of the
polytope $\mathcal{M}(2,p)$, contrasting sharply with the case $n\geq 3$, $%
p>n$ where $UNI(p)$ is an extreme point of the non convex set $\mathcal{M}%
(n,p)$: Theorem 2.\smallskip

\textbf{Proof}\textit{.} Fix $\lambda \in \mathcal{G}(2,p)$ and symmetric.
Rewrite (\ref{4}) as $\lambda \cdot u^{\ast }=\widetilde{\lambda }\cdot
u^{\ast }$ for all $u^{\ast }$, which by the identity (\ref{15}) means $%
\lambda \cdot u^{\ast }=-\lambda \cdot (-u)^{\ast }$ for all $u^{\ast }$.
The latter is property (\ref{16}) for $n=2$, so the maximality of $\lambda $
follows as in the above proof of statement $i)$ Proposition 1.

To prove the converse statement pick $\lambda \in \mathcal{G}(2,p)$, which
means $\widetilde{\lambda }\vdash \lambda $ (see property (\ref{12}) in the
previous section). As the dominance relation is preserved by convex
combinations we have $\frac{1}{2}(\widetilde{\lambda }+\lambda )\vdash
\lambda $ where $\frac{1}{2}(\widetilde{\lambda }+\lambda )$ is symmetric:
thus $\lambda $ is dominated if it is not symmetric. $\blacksquare
\smallskip $

The protocols implementing the vertices of $\mathcal{M}(2,p)$ combine in a
simple way the veto and random dictator ideas explained in section 1. Asking
one randomly chosen agent to select a pure outcome implements $\lambda ^{1}=(%
\frac{1}{2},0,\cdots ,0,\frac{1}{2})$, which we call the \textit{random
dictator} guarantee and denote $RD(2,p)$. The guarantee $VT(2,p)$ is $%
\lambda =(0,\frac{1}{p-2},\cdots ,\frac{1}{p-2},0)$ implemented by giving
one veto token per person, then selecting one of the remaining outcomes with
uniform probability (as in the example of section 1). It is maximal, though
not a vertex of $\mathcal{M}(2,p)$ except if $p=3$ or $4$.

To implement $\lambda ^{2}=(0,\frac{1}{2},0,\cdots ,0,\frac{1}{2},0)$ we ask
first each agent to veto one outcome, after which we pick a random dictator
between the remaining ($p-2$ or $p-1$) outcomes: we write this guarantee as $%
{\large VT}\times RD(2,p)$. And so on: $VT^{t-1}\times RD(2,p)$ is the
guarantee $\lambda ^{t}$: let each agent veto $t-1$ pure outcomes, then pick
a random dictator for the remaining ones. If $p$ is odd the guarantee $%
\lambda ^{\frac{p+1}{2}}$ denoted $VT^{\frac{p-1}{2}}(2,p)$ simply gives $%
\frac{p-1}{2}$ veto tokens to each agent.

\subsection{A duality operation preserving $\mathcal{M}(n,p)$}

Although the results in this subsection apply for all $n,p$, they are only
useful if $3\leq n<p$, the only case not covered by Propositions 1 and 2: we
maintain this assumption from now on. Besides the uniform rule $UNI(p)$ the
two most basic guarantees come from one round of veto or of random dictator:%
\begin{equation*}
VT(n,p)\rightarrow (0,\frac{1}{p-n},\cdots ,\frac{1}{p-n},\overset{n-1}{%
\overbrace{0,\cdots ,0}})
\end{equation*}%
\begin{equation*}
RD(n,p)\rightarrow (\overset{{\large n-1}}{\overbrace{\frac{1}{n},\cdots ,%
\frac{1}{n}}},0,\cdots ,0,\frac{1}{n})
\end{equation*}

Given a lottery $\lambda $ in $\Delta (p)$ different from $\lambda ^{uni}$,
the \textit{radius to }$\lambda $ is the interval of the half-line from $%
\lambda ^{uni}$ toward $\lambda $ contained in $\Delta (p)$ (its other end
is on the boundary $\partial \Delta (p)$), i. e. all lotteries of the form $%
\lambda ^{uni}+\alpha (\lambda -\lambda ^{uni})$ for some $\alpha \geq 0$.
The \textit{anti-radius from }$\widetilde{\lambda} $ is the interval in $%
\Delta (p)$ of the half-line from $\lambda ^{uni}$ away from $\widetilde{%
\lambda} $, i. e., the set of all lotteries of the form $\lambda
^{uni}+\alpha (\lambda ^{uni}-\widetilde{\lambda} )$ for some $\alpha \geq 0$%
.

If $\lambda $ is a boundary lottery its dual $\lambda ^{\bigstar }$ is the
end point of the anti-radius from\textit{\ }$\widetilde{\lambda }$%
\begin{equation}
\lambda ^{\bigstar }=(1+\alpha )\lambda ^{uni}-\alpha \widetilde{\lambda }%
\text{ where }\alpha =\frac{1}{p\cdot \max_{1\leq k\leq p}\lambda _{k}-1}
\label{28}
\end{equation}%
(where $\max_{1\leq k\leq p}\lambda _{k}>\frac{1}{p}$ because $\lambda \in
\partial \Delta (p)$). Keeping in mind that $\min_{1\leq k\leq p}\lambda
_{k}=0$ it is easy to check the identity $(\lambda ^{\bigstar })^{\bigstar
}=\lambda $. For non boundary lotteries we extend this definition linearly
on the radius to $\lambda $%
\begin{equation*}
\{\mu \in \partial \Delta (p)\text{ and }\lambda =\delta \lambda
^{uni}+(1-\delta )\mu \}\Longrightarrow \lambda ^{\bigstar }=\delta \lambda
^{uni}+(1-\delta )\mu ^{\bigstar }.
\end{equation*}
so that $\lambda \rightarrow \lambda ^{\bigstar }$ is a proper duality in $%
\Delta (p)$.

The uniform lottery is the only self-dual one, while $VT(n,p)$ and $RD(n,p)$
are dual of each other.\textbf{\smallskip }

\textbf{Proposition 3}

\noindent $i)$ \textit{If }$\lambda \ne \lambda^{uni}$ \textit{is a maximal
guarantee, the radius to }$\lambda $\textit{\ and the anti-radius from }$%
\widetilde{\lambda }$\textit{\ (the symmetric of }$\lambda $\textit{\ w.r.t.
the middle rank)} \textit{are contained in} $\mathcal{M}(n,p)$.

\noindent $ii)$ \textit{The duality operation }$\lambda \rightarrow \lambda
^{\bigstar }$\textit{\ in }$\Delta (p)$ \textit{preserves maximal lotteries:}%
\begin{equation*}
\lbrack \mathcal{M}(n,p)]^{\bigstar }=\mathcal{M}(n,p)
\end{equation*}

Note that the statement in Proposition 3 holds if we replace $\mathcal{M}%
(n,p)$ by $\mathcal{G}(n,p)$: the radius to a feasible guarantee, and the
anti-radius from its symmetric are feasible as well; duality preserves
feasibility. This follows at once from the Corollary to Lemma 1, the
identity (\ref{15}) and the definition (\ref{28}).

The proof for $\mathcal{M}(n,p)$ is much harder: we need a technical result
characterising $\mathcal{M}(n,p)$ in $\mathcal{G}(n,p)$ by its position w.
r. t. the polar cone of $\mathcal{G}(n,p)$. Notation: we write $G^{\ominus }$
for the polar cone of $G\subset 
\mathbb{R}
^{p}$: $G^{\ominus }=\{z\in 
\mathbb{R}
^{p}|\forall y\in G:z\cdot y\leq 0\}$.\smallskip

\textbf{Lemma 2 }\textit{The guarantee }$\lambda \in \mathcal{G}(n,p)$%
\textit{\ is maximal if and only if there exists a vector }$z\in \mathcal{G}%
(n,p)^{\ominus }$ \textit{s. t. }$\sum_{k=1}^{p}z_{k}=0$,\textit{\ }$%
z_{1}<z_{2}<\cdots <z_{p}$\textit{\ and} $\lambda \cdot z=0$.$\smallskip $

\textbf{Proof of \textquotedblleft if\textquotedblright }. Fix $\lambda $ in 
$\mathcal{G}(n,p)$ and $z$ in $\mathcal{G}(n,p)^{\ominus }$ as in the
statement, and suppose $\lambda $ is dominated by $\mu $. As the coordinates
of $z$ increase strictly, $\mu \vdash \lambda $ and $\mu \neq \lambda $
imply $\lambda \cdot z<\mu \cdot z$. Now feasibility of $\mu $ and $z\in 
\mathcal{G}(n,p)^{\ominus }$ give $\mu \cdot z\leq 0$. This contradicts the
assumption $\lambda \cdot z=0$. $\blacksquare $ \smallskip

Note that the condition $\sum_{k=1}^{p}z_{k}=0$ was not used, therefore
Lemma 2 remains valid without this condition. But the condition makes the
\textquotedblleft only if\textquotedblright\ part stronger. The long proof
of this direction is given in the Appendix.\smallskip

\textbf{Proof of Proposition 3.}

\noindent \textit{Statement }$i)$

We fix $\lambda \in \mathcal{M}(n,p)$ and $z\in \mathcal{G}(n,p)^{\ominus }$
as in Lemma 2. Consider first a lottery $\mu =\lambda ^{uni}+\alpha (\lambda
-\lambda ^{uni})$ in the radius to $\lambda $. That $\mu $ is feasible as
well ($\mu \in \mathcal{G}(n,p)$) is clear by checking property (\ref{2}) in
section 3. For maximality we use $\sum_{k=1}^p z_k=0$ and $\lambda \cdot z=0$
to compute $\mu \cdot z=(1-\alpha )\lambda^{uni} \cdot z +\alpha \lambda
\cdot z=0$ and conclude $\mu \in \mathcal{M}(n,p)$ by Lemma 2 again.

Still fixing $\lambda \in \mathcal{M}(n,p)$ and $z$, we pick next a lottery $%
\mu =\lambda ^{uni}+\alpha (\lambda ^{uni}-\widetilde{\lambda })$ in the
anti radius from $\widetilde{\lambda }$. For feasibility we check property (%
\ref{2}) at an arbitrary profile $(u_{i})_{i=1}^{n}$ s. t. $%
\sum_{1}^{n}u_{i}=0$. Compute%
\begin{equation*}
\sum_{1}^{n}\mu \cdot u_{i}^{\ast }=(1+\alpha )\sum_{1}^{n}\lambda
^{uni}\cdot u_{i}-\alpha \sum_{1}^{n}\widetilde{\lambda }\cdot u_{i}^{\ast
}=\alpha \sum_{1}^{n}\lambda \cdot (-u_{i})^{\ast }\leq 0
\end{equation*}%
where the last equality is the identity (\ref{15}), and the inequality is
from property (\ref{2}) for $\lambda$.

The argument just made shows that for any $\xi \in \mathcal{G}(n,p)$ the
lottery $(1+\alpha )\lambda ^{uni}-\alpha \widetilde{\xi }$ is feasible as
well, in particular%
\begin{equation*}
0\geq ((1+\alpha)\lambda^{uni} - \alpha \widetilde{\xi}) \cdot z=-\alpha 
\widetilde{\xi }\cdot z
\end{equation*}%
where the equality uses $\sum_{k=1}^p z_k=0$. Writing 
\begin{equation*}
w=(-z_p, -z_{p-1}, \ldots, -z_1)
\end{equation*}
and using the identity (\ref{15}), we conclude that $\xi \cdot w \le 0$.
Thus $w$ is in $\mathcal{G}(n,p)^{\ominus }$ too, and it satisfies the
requirements in Lemma 2 with respect to $\mu$: $\mu \cdot w = -\alpha 
\widetilde{\lambda} \cdot w = \alpha \lambda \cdot z = 0$, which proves the
maximality of $\mu$.\smallskip

\noindent \textit{Statement }$ii)$ follows from statement $i)$ and the
definition of the duality operation. $\blacksquare $

\subsection{Canonical guarantees}

We write the largest coordinate of a lottery as $\lambda _{+}=\max_{1\leq
k\leq p}\lambda _{k}$. We see from (\ref{28}) that the dual $\lambda
^{\bigstar }$ of the boundary lottery $\lambda $ is

\begin{equation}
\lambda _{k}^{\bigstar }=\frac{1}{p\lambda _{+}-1}(\lambda _{+}-\widetilde{%
\lambda }_{k})\text{ for }1\leq k\leq p  \label{29}
\end{equation}%
(where $\widetilde{\lambda }_{k}=\lambda _{p+1-k}$)\smallskip

\textbf{Definition 5} \textit{Composition by }$VT$\textit{\ and }$RD$

\noindent \textit{For any }$\lambda \in \Delta (p)$\textit{\ the lottery }$%
VT\otimes \lambda \in \Delta (p+n)$\textit{\ obtains by inserting }$\lambda $%
\textit{\ between one zero in rank 1 and }$n-1$\textit{\ zeros after rank }$%
p+1$\textit{:}%
\begin{equation}
VT\otimes \lambda =(0,\lambda ,\overset{n-1}{\overbrace{0,\cdots ,0}})
\label{30}
\end{equation}%
\textit{If }$\lambda \in \partial \Delta (p)$\textit{\ the lottery }$%
RD\otimes \lambda \in \partial \Delta (p+n)$ \textit{obtains by filling
uniformly }$n-1$\textit{\ ranks before }$\lambda $ \textit{and one after as
follows:}%
\begin{equation}
RD\otimes \lambda =(\overset{{\large n-1}}{\overbrace{\frac{\lambda _{+}}{%
n\lambda _{+}+1},\cdots ,\frac{\lambda _{+}}{n\lambda _{+}+1}}},\frac{1}{%
n\lambda _{+}+1}\cdot \lambda ,\frac{\lambda _{+}}{n\lambda _{+}+1})
\label{31}
\end{equation}%
\textit{For any }$\lambda \in \Delta (p)$\textit{\ the lottery }$RD\otimes
\lambda \in \Delta (p+n)$ \textit{is given by }%
\begin{equation}
RD\otimes \lambda =[VT\otimes \lambda ^{\bigstar }]^{\bigstar }  \label{32}
\end{equation}

If $\lambda \in \partial \Delta (p)$ we must check that the two definitions (%
\ref{31}) and (\ref{32}) coincide. Write $\mu $ for the boundary lottery on
the right-hand side of equation (\ref{31}): applying (\ref{29}) and $\mu
_{+}=\frac{\lambda _{+}}{n\lambda _{+}+1}$ we get 
\begin{equation*}
\mu ^{\bigstar }=\frac{1}{(p+n)\frac{\lambda _{+}}{n\lambda _{+}+1}-1}(\frac{%
\lambda _{+}}{n\lambda _{+}+1}\boldsymbol{1}-\widetilde{\mu }) \quad \quad
\quad \quad \quad \quad \quad \quad
\end{equation*}%
\begin{equation*}
=\frac{1}{p\lambda _{+}-1}(\lambda _{+}\boldsymbol{1}-(\lambda_+,\widetilde{%
\lambda},\overset{{\large n-1}}{\overbrace{\lambda_+,\ldots,\lambda_+}}%
))=VT\otimes \lambda ^{\bigstar }
\end{equation*}
as desired.

Note that Definition 5 implies in particular $VT\otimes UNI(p)=VT(n,n+p)$
and $RD\otimes UNI(p)=RD(n,n+p)$.\smallskip

\textbf{Lemma 3}

\noindent $i)$ \textit{The guarantees }$VT(n,p)$ \textit{and} $RD(n,p)$%
\textit{\ are maximal.}

\noindent $ii)$\textit{\ The composition of guarantees} \textit{by }$VT$%
\textit{\ and }$RD$\textit{\ respects their feasibility and maximality. For
any }$\lambda \in \Delta (p)$%
\begin{equation*}
\lambda \in \mathcal{M}(n,p)\Longrightarrow VT\otimes \lambda ,RD\otimes
\lambda \in \mathcal{M}(n,p+n)\text{ }
\end{equation*}%
\textit{and the same statement holds by replacing }$\mathcal{M}(n,p)$ 
\textit{by} $\mathcal{G}(n,p)$\textit{\ and} $\mathcal{M}(n,p+n)$ \textit{by}
$\mathcal{G}(n,p+n).\smallskip $

For the proof we need a second characterisation of maximal guarantees; the
proof, much easier than that of Lemma 2, is also in the Appendix.\smallskip

\textbf{Lemma 4} \textit{The guarantee }$\lambda \in \mathcal{G}(n,p)$%
\textit{\ is maximal if and only if for all }$k\in \lbrack p-1]$\textit{\
there exists a preference profile }$\pi $\textit{\ such that, for any
lottery }${\large \ell }$\textit{\ implementing }$\lambda $\textit{\ at }$%
\pi $\textit{\ (Definition 1) we have}%
\begin{equation}
\max_{i\in \lbrack n]}[\ell ^{\ast i}]_{1}^{k}=[\lambda ]_{1}^{k}  \label{13}
\end{equation}

\textbf{Proof of Lemma 3}

\noindent Statement $i)$\textit{\ }The proof that $VT(n,p)$ is maximal, done
in section 1 for $n=3,p=6$, is an application of Lemma 4. Its generalisation
is straightforward. Then its dual $RD(n,p)$ is maximal by Proposition 3.

\noindent \textit{Statement }$ii)$ Fixing $\lambda \in \mathcal{G}(n,p)$ we
implement $VT\otimes \lambda $ as follows: ask agents to report their worst
outcome, eliminate $n$ outcomes containing all the reported ones, then
implement $\lambda $ over the remaining $p$\ outcomes. The latter are ranked
weakly higher than $2,\cdots ,p+1$ for each agent, so we conclude that $%
VT\otimes \lambda $ is feasible.

If now $\lambda \in \mathcal{M}(n,p)$, we fix an index $k\in \lbrack p-1]$
and an $(n,p)$-profile $\pi $ ensuring property (\ref{13}) as in the
premises of Lemma 4. We construct the following $(n,p+n)$ profile $\theta $%
\begin{equation}
\begin{array}{cccccc}
\prec _{1} & a_{1} & \overset{p}{\overbrace{\pi }} & a_{2} & \cdots & a_{n}
\\ 
\cdots & \cdots & \pi & \cdots &  & \cdots \\ 
\prec _{n} & a_{n} & \pi & a_{1} & \cdots & a_{n-1}%
\end{array}
\label{33}
\end{equation}%
where the initial profile $\pi $ on $p$ outcomes occupies the ranks $2$ to $%
p+1$, while the preferences over the $n$ other outcomes are cyclical. If a
lottery $\ell $ implements $VT\otimes \lambda $ at $\theta $ it can put no
weight on any $a_{i}$ outcome because $(VT\otimes \lambda )_{1}=0$,
therefore the restriction of $\ell$ to the outcomes of $\pi$ implements $%
\lambda$ at $\pi $, so property (\ref{13}) holds for ranks $2$ to $p+1$ as
well as for the first one and the last $n-1$ ones.

That $RD\otimes \lambda $ is feasible, resp. maximal if $\lambda $ is
follows from the duality relation (\ref{32}) and the fact that duality
respects maximality and feasibility (Proposition 3). Here is for
completeness the protocol implementing $RD\otimes \lambda $ if $\lambda $ is
a boundary feasible lottery: agents report their best outcome, then we pick $%
n$ outcomes containing all reports; with probability $\frac{n\lambda _{+}}{%
n\lambda _{+}+1}$ we choose one of those uniformly, and with probability $%
\frac{1}{n\lambda _{+}+1}$ we implement $\lambda $ among the remaining $p$
outcomes. $\blacksquare \smallskip $

\textbf{Definition 6} \textit{Canonical guarantees}

\noindent \textit{Fix }$n,p,3\leq n<p$\textit{, s. t. }$d=\lfloor \frac{p-1}{%
n}\rfloor $ and\textit{\ }$p=dn+q$ for some $q=1,\cdots ,n$\textit{. Each
sequence }$\Gamma =(\Gamma ^{t})_{t=1}^{h}$\textit{\ in }$\{VT,RD\}$\textit{%
\ of length }$h,h\leq d$\textit{, defines a canonical guarantee }$\Gamma
^{1}\otimes \Gamma ^{2}\otimes \cdots \otimes \Gamma ^{h}$ \textit{by
iterating the composition operation, i.e., }%
\begin{equation*}
\Gamma ^{1}\otimes \Gamma ^{2}\otimes \cdots \otimes \Gamma ^{h}=\Gamma
^{1}\otimes (\Gamma ^{2}\otimes (\cdots \otimes \Gamma ^{h}) \cdots )
\end{equation*}%
\textit{where } $\Gamma^h$ \textit{\ acts on } $(d-h+1)n + q$ \textit{\
outcomes, } $\Gamma^{h-1} \otimes \Gamma^h$ \textit{\ acts on } $(d-h+2)n +
q $ \textit{\ outcomes, etc. } \textit{We write their set as} $\mathcal{C}%
(n,p) $\textit{, of cardinality} $2^{d+1}-2$.\smallskip

By Lemma 3 and the fact that the composition by each $\Gamma ^{t}$ adds $n$
outcomes to the previous ones, all canonical lotteries are maximal. By
duality (\ref{32}), canonical lotteries come in dual pairs: exchanging $VT$
and $RD$ in each term of the sequence $\Gamma $ produces the dual lottery.

An important observation is that each $\lambda \in \mathcal{C}(n,p)$ is
uniform on its support, therefore determined by this non full support. This
implies that it is a vertex of $\mathcal{G}(n,p)$ (the proof mimicks that of
statement $iii)$ in Proposition 1), hence also a vertex of $\mathcal{M}(n,p)$%
.

We give some examples.

If $d=1$ ($p\leq 2n$) $VT(n,p)$ and $RD(n,p)$ are the only canonical
guarantees.

Constant sequences: the composition of $h$ veto steps, or of $h$ random
dictator steps, gives dual lotteries of a similar shape: their support is at
the extreme ranks or in the center:

\begin{equation*}
\overset{h}{\overbrace{VT\otimes \cdots \otimes VT}}=(\overset{h}{\overbrace{%
0,\cdots ,0}},\frac{1}{p-nh},\cdots ,\frac{1}{p-nh},\overset{(n-1)h}{%
\overbrace{0,\cdots ,0}})
\end{equation*}%
\begin{equation*}
\overset{h}{\overbrace{RD\otimes \cdots \otimes RD}}=(\overset{(n-1)h}{%
\overbrace{\frac{1}{nh},\cdots ,\frac{1}{nh}}},0,\cdots ,0,\overset{h}{%
\overbrace{\frac{1}{nh},\cdots ,\frac{1}{nh}}})
\end{equation*}

A simple protocol for the former gives $h$ veto tokens to each agent, then
randomises uniformly between the remaining outcomes, even if there are more
than $p-nh$ of those (which will only improve the guaranteed welfare). To
implement the latter we elicit from each agent her $h$ top outcomes, then
randomise uniformly between any $nh$ outcomes containing all reported tops,
adding arbitrary outcomes if the reported ones are fewer than $nh$. The last
instruction is important: ignoring it could result in giving too much weight
to someone's worst outcomes (as illustrated in the example of section
1).\smallskip

For $d=2$ we have six canonical guarantees, four from the constant sequences
and a dual pair from $(VT,RD)$ and $(RD,VT)$. For instance in $\mathcal{C}%
(3,7)$:%
\begin{equation*}
VT\otimes RD=(0,\frac{1}{3},\frac{1}{3},0,\frac{1}{3},0,0)\text{ ; }%
RD\otimes VT=(\frac{1}{4},\frac{1}{4},0,\frac{1}{4},0,0,\frac{1}{4})
\end{equation*}%
The protocol for $RD\otimes VT$ selects three outcomes containing the top
ones of each agent; then with probability $3/4$ it picks one of those
uniformly, and with probability $1/4$ plays $VT(3,4)$ among the remaining
outcomes.

Our final example is in $\mathcal{C}(3,11)$ where $d=3$ and we have three
pairs of non constant sequences of length three, for instance:%
\begin{equation*}
(RD,VT,VT)\rightarrow \lambda =(\frac{1}{5},\frac{1}{5},0,0,\frac{1}{5},%
\frac{1}{5},0,0,0,0,\frac{1}{5})
\end{equation*}%
\begin{equation*}
(RD,VT,RD)\rightarrow \lambda =(\frac{1}{6},\frac{1}{6},0,\frac{1}{6},\frac{1%
}{6},0,0,\frac{1}{6},0,0,\frac{1}{6})
\end{equation*}

\section{General results with three or more agents}

\subsection{Maximal guarantees for $3\leq n<p\leq 2n$}

If $d=1$ we have only two canonical guarantees $VT(n,p)$ and $RD(n,p)$ and
by Proposition 3 any convex combination of $UNI(p)$ with one of these two is
also maximal. It turns out that, for the most part, this exhausts all
maximal guarantees.\smallskip

\textbf{Theorem 1}

\noindent $i)$ \textit{For any }$n,p$ \textit{s. t. }$3\leq n<p$\textit{\
let }$\lambda ^{vt},\lambda ^{rd},\lambda ^{uni}$ be the guarantees from $%
VT(n,p),RD(n,p)$ and $UNI(p)$. \textit{Then\ }%
\begin{equation}
\lbrack \lambda ^{uni},\lambda ^{vt}]\cup \lbrack \lambda ^{uni},\lambda
^{rd}]\subset \mathcal{M}(n,p)  \label{35}
\end{equation}%
$ii)$\textit{\ This is an equality if }$p\leq 2n-2$\textit{\ and if }$p=2n$%
\textit{\ except when }$(n,p)=(4,8)$\textit{\ or }$(5,10)$.\textit{%
\smallskip }

The proof is in the Appendix.

Our next result explains why additional maximal guarantees appear in the
cases excluded by statement $ii)$ above and describes the full set $\mathcal{%
M}(n,p)$ in two such cases.\smallskip

\textbf{Proposition 4}

\noindent $i)$ \textit{If} $p=2n-1$\textit{\ and if }$(n,p)=(4,8)$\textit{\
or }$(5,10)$\textit{, the inclusion (\ref{35}) is strict.}

\noindent $ii)$\textit{\ For }$n=3,$ $p=5$ \textit{there are two dual pairs
of maximal guarantees on the boundary of }$\Delta (5)$\textit{: }$%
VT(3,5),RD(3,5)$\textit{\ and the pair}%
\begin{equation*}
\lambda =(\frac{1}{2},0,0,\frac{1}{2},0)\text{ ; }\lambda ^{\bigstar }=(%
\frac{1}{3},0,\frac{1}{3},\frac{1}{3},0)
\end{equation*}%
\textit{The set }$\mathcal{M}(3,5)$ \textit{is the union of the four
intervals joining }$UNI(5)$\textit{\ to these guarantees.}

\noindent $iii)$\textit{\ For }$n=4,$ $p=7$ \textit{there are three dual
pairs of maximal guarantees on the boundary of }$\Delta (7)$\textit{: }$%
VT(4,7),RD(4,7)$\textit{\ and the two pairs}%
\begin{equation*}
\lambda =(\frac{1}{2},0,0,0,\frac{1}{2},0,0)\text{ ; }\lambda ^{\bigstar }=(%
\frac{1}{5},\frac{1}{5},0,\frac{1}{5},\frac{1}{5},\frac{1}{5},0)
\end{equation*}%
\begin{equation*}
\mu =(\frac{1}{3},\frac{1}{9},\frac{2}{9},0,0,\frac{1}{3},0)\text{ ; }\mu
^{\bigstar }=(\frac{1}{4},0,\frac{1}{4},\frac{1}{4},\frac{1}{12},\frac{1}{6}%
,0)
\end{equation*}
\textit{The set} $\mathcal{M}(4,7)$ \textit{is the union of the six
intervals joining} $UNI(7)$ \textit{to these guarantees.}

\textbf{Proof}

\noindent \textit{Statement }$i)$. Assume $p=2n-1$. At any profile we can
choose a set of $n-1$ outcomes meeting (containing at least one of) the top
two outcomes of each agent. A uniform lottery over these outcomes guarantees
to every agent a probability of at least $\frac{1}{n-1}$ for his top two
outcomes. Hence there must be a maximal guarantee that does that, but
neither $UNI(p),VT(n,p)$ nor $RD(n,p)$ does this, hence neither does a
convex combination of these.

For $(4,8)$ one checks easily that we can always choose a triple of outcomes
meeting the top three outcomes of each agent in \textit{at most} one
element. A uniform lottery over the complement of that triple guarantees to
every agent at least $\frac{2}{5}$ for his top three outcomes, and the
argument is completed as above. For $(5,10)$, a simple case check shows that
we can choose a triple of outcomes meeting the top three outcomes of each
agent. A uniform lottery over them guarantees to every agent at least $\frac{%
1}{3}$ for his top three outcomes, and the argument is completed as above.

\noindent \textit{Statements }$ii)$ and $iii)$. The protocols implementing $%
\lambda $ and $\lambda ^{\bigstar }$ in each case follow the same logic as
above. For $\lambda $ we can always pick two outcomes $x,y$ meeting the top
two (when $(n,p)=(3,5)$) or three (when $(n,p)=(4,7)$) of any agent, then we
draw $x$ and $y$ each with probability $\frac{1}{2} $. For $\lambda
^{\bigstar }$ we can always pick two outcomes $x,y$ such that the worst two
(when $(n,p)=(3,5)$) or three (when $(n,p)=(4,7)$) of any agent contain at
least one of them, then we randomise uniformly over the remaining outcomes.

We omit for brevity the tedious arguments, available upon request from the
authors, showing that these guarantees, as well as $\mu $ and $\mu
^{\bigstar }$ are maximal, and generate the entire sets $\mathcal{M}(3,5)$
and $\mathcal{M}(4,7)$. $\blacksquare \smallskip $

Note that, in particular, Theorem 1 and Proposition 4 give a full
description of maximal guarantees whenever $3\leq n<p\leq n+3$.

\subsection{Maximal guarantees for $3\leq n<p$}

For higher values of $d=\lfloor \frac{p-1}{n}\rfloor $ we know only a few
general facts about the structure of $\mathcal{M}(n,p)$. Lemma 2 in
Proposition 3 provides our best clue. For any $z\in \mathcal{G}%
(n,p)^{\ominus }$ such that $\mathcal{G}(n,p)$ intersects the hyperplane $%
H=\{y|z\cdot y=0\}$, the intersection $H\cap \mathcal{G}(n,p)$ is a face of $%
\mathcal{G}(n,p)$, in particular a polytope. The Lemma tells us that such a
face defined by a vector $z$ with increasing coordinates is a subset of $%
\mathcal{M}(n,p)$, and that all maximal guarantees obtain for some $z$. This
proves the following.\smallskip

\textbf{Proposition 5} \textit{For } $3 \le n < p$, \textit{\ the set} $%
\mathcal{M}(n,p)$ \textit{is a finite union of faces of the polytope }$%
\mathcal{G}(n,p)$, \textit{\ each having} $UNI(p)$ \textit{as a vertex.}%
\smallskip

Our second main result identifies a large subset of $\mathcal{M}(n,p)$
constructed from the canonical guarantees.\smallskip

\textbf{Theorem 2 }\textit{Fix }$n,p$ \textit{s. t}. $3\leq n<p,$\textit{\ }$%
d=\lfloor \frac{p-1}{n}\rfloor$\textit{.}

\noindent \textit{For each sequence }$\Gamma $ \textit{of length }$d$ 
\textit{in }$\{VT,RD\}$\textit{, the canonical guarantees from the }$d$%
\textit{\ initial subsequences\footnote{%
I. e., the guarantees $\Gamma ^{1},\Gamma ^{1}\otimes \Gamma ^{2},\Gamma
^{1}\otimes \Gamma ^{2}\otimes \Gamma ^{3}$, etc..} of }$\Gamma $\textit{,
plus the uniform guarantee, are the vertices of a simplex of dimension }$d$%
\textit{\ contained in} $\mathcal{M}(n,p)$.$\smallskip $

The proof is in the Appendix.

The simplest example not covered in Theorem 1 is $n=3,p=7$, so $d=2$.
Theorem 2 describes four triangles of maximal guarantees coming in dual
pairs. The uniform lottery is always a vertex and the other two vertices are
canonical guarantees:%
\begin{equation*}
\begin{array}{ccc}
\text{sequence} & \text{vertex 1} & \text{vertex 2} \\ 
VT,VT & (0,\frac{1}{4},\frac{1}{4},\frac{1}{4},\frac{1}{4},0,0) & 
(0,0,1,0,0,0,0) \\ 
RD,RD & (\frac{1}{3},\frac{1}{3},0,0,0,0,\frac{1}{3}) & (\frac{1}{6},\frac{1%
}{6},\frac{1}{6},\frac{1}{6},0,\frac{1}{6},\frac{1}{6}) \\ 
VT,RD & (0,\frac{1}{4},\frac{1}{4},\frac{1}{4},\frac{1}{4},0,0) & (0,\frac{1%
}{3},\frac{1}{3},0,\frac{1}{3},0,0) \\ 
RD,VT & (\frac{1}{3},\frac{1}{3},0,0,0,0,\frac{1}{3}) & (\frac{1}{4},\frac{1%
}{4},0,\frac{1}{4},0,0,\frac{1}{4})%
\end{array}%
\end{equation*}%
where the dual pairs are the top two and the bottom two rows. In addition to
these four triangles, the maximal set $\mathcal{M}(3,7)$ also contains two
intervals, joining $UNI(7)$ to each of the following two dual non canonical
guarantees: 
\begin{equation*}
\lambda =(\frac{1}{3},0,0,\frac{1}{3},\frac{1}{3},0,0)\text{ ; }\lambda
^{\bigstar }=(\frac{1}{4},\frac{1}{4},0,0,\frac{1}{4},\frac{1}{4},0)
\end{equation*}

In general, we keep in mind that many more guarantees than the ones
described in Theorem 2 are maximal. Pick any non canonical guarantee $%
\lambda $ in $\mathcal{M}(n,p)\cap \partial \Delta (p)$, for instance those
described in Proposition 4 or in the previous paragraph: by Lemma 3
successive compositions of $\lambda $ with $VT$ and/or $RD$ generate, for
any $h\geq 1$, $2^{h}$ non canonical maximal guarantees in $\mathcal{M}%
(n,p+hn)\cap \partial \Delta (p+hn)$.

\section{Concluding comments}

The set $\mathcal{M}(n,p)$ remains simple if $n=2$ and/or $p\leq 2n$
(Proposition 2 and Theorem 1), but its combinatorial/geometric structure
becomes complicated, perhaps severely, as $\frac{p}{n}$ increases while $%
n\geq 3$. Questions that remain open for further investigation include:

\begin{itemize}
\item can we get new maximal guarantees by other convex combinations of
canonical guarantees than those described in Theorem 2? We conjecture the
answer is No.

\item what is the maximal dimension of a simplicial component of $\mathcal{M}%
(n,p)$? We conjecture it is $d=\lfloor \frac{p-1}{n}\rfloor $.

\item can we evaluate the number of such components?
\end{itemize}

The early voting by veto literature stresses the guarantees it offers to
coalitions of like-minded voters (\cite{Mue}). We could similarly define,
given $n$ and $p$, a guarantee for each size of a coalition, and try to link
our modeling approach to the design of voting rules where the strategic
formation of coalitions promotes stability (\cite{Mou5}, \cite{Mou6},\cite%
{Pel}).

Even for the individual guarantees discussed here, it may be possible to
connect a maximal guarantee with the "best" game form(s) to implement it,
where best may refer to simplicity, or to strategic or normative properties.

\section{Appendix}

\subsection{Proof of Lemma 2}

We prove the only if statement: for any $\lambda \in \mathcal{M}(n,p)$ we
can find a vector $z$ as in Lemma 2.

Consider the following cone $W$ in $%
\mathbb{R}
^{p}$:%
\begin{equation}
W={\large \{}z=\sum_{i=1}^{n}u_{i}^{\ast }{\large |}\text{ for some }%
U=(u_{i})_{i=1}^{n}\text{ s.t. }\sum_{i=1}^{n}u_{i}=0{\large \}}  \label{22}
\end{equation}

By its characteristic property (\ref{2}) $\mathcal{G}(n,p)$ is the
intersection of $W^{\ominus }$ with $\Delta (p)$, therefore $\mathcal{G}%
(n,p)^{\ominus}$ is the Minkowski sum of $\overleftrightarrow{W}$ and $%
\mathbb{R}
_{-}^{p}$, where $\overleftrightarrow{W}$ is the convex hull of $W$.
Moreover the identity $\sum_{(i,k)\in \lbrack n]\times \lbrack
p]}u_{ik}^{\ast }=\sum_{(i,a)\in \lbrack n]\times A}u_{ia}$ implies $%
\sum_{k=1}^{p}z_{k}=0$ in $W$, therefore $\overleftrightarrow{W}=\mathcal{G}%
(n,p)^{\ominus }\cap \{z|\sum_{k=1}^{p}z_{k}=0\}$.

We fix now a maximal guarantee $\lambda $ and define the sub-cone $Z$ of $%
\overleftrightarrow{W}$:%
\begin{equation*}
Z={\large \{}z\in \mathcal{G}(n,p)^{\ominus }|\sum_{k=1}^{p}z_{k}=0\mathit{\ 
}\text{and }\lambda \cdot z=0{\large \}}
\end{equation*}

This cone is convex, and every element of $Z$ satisfies $z_{1}\leq z_{2}\leq
\cdots \leq z_{p}$, because these inequalities hold in $W$. To prove that $Z$
contains some $z$ s. t. $z_{1}<z_{2}<\cdots <z_{p}$, we choose in $Z$ one $%
\widehat{z}$ in which the number of equalities between consecutive
coordinates of $\widehat{z}$ is as small as possible. If there is no
equality we are done. Otherwise, assume that the first equality is $\widehat{%
z}_{k}=\widehat{z}_{k+1}$. We will show the existence of some $z\in W$ s. t. 
$z_{k}<z_{k+1}$ and $\lambda \cdot z=0$: this leads to a contradiction
because $\widehat{z}+z\in Z$\textbf{\ }has fewer equalities than $\widehat{z}
$. Consider two cases.

\textbf{Case 1} $\lambda _{k}>0$

We proceed by contradiction and assume that if $z\in W$ and $z_{k}<z_{k+1}$,
then $\lambda \cdot z<0$.

Call ${\large \Pi }$ the set of profiles $U=(u_{i})_{i=1}^{n}$ such that%
\begin{equation}
\sum_{i=1}^{n}u_{i}=0\text{ and }u_{1k}^{\ast}=0\text{ , }u_{1,k+1}^{\ast}=1
\label{20}
\end{equation}

The corresponding vector $z=\sum_{i=1}^{n}u_{i}^{\ast }$ is in $W$ therefore 
$\sum_{i=1}^{n}\lambda \cdot u_{i}^{\ast }<0$ for all $U\in {\large \Pi }$.
We show next, again by contradiction, that the supremum of $%
\sum_{i=1}^{n}\lambda \cdot u_{i}^{\ast }$ over ${\large \Pi }$ cannot be
zero.

If it is, there is a sequence $U^{s}$ in ${\large \Pi }$ s. t. the sequence $%
\sum_{i=1}^{n}\lambda \cdot u_{i}^{s\ast }$ converges to zero. By taking
subsequences, we can make sure that for each $i$, the way each $u_{i}^{s}$
orders the outcomes in $A$ does not depend on $s$ (but depends on $i$). Then
for each $i$ there is a lottery $\lambda ^{i}$ on $A$, its coordinates a
permutation of those of $\lambda $, s.t. $\lambda \cdot u_{i}^{s\ast
}=\lambda ^{i}\cdot u_{i}^{s}$ for all $s$.

Consider the polyhedron $Q$ of $n\times p$ matrices $X=[x_{i}^{a}]_{i\in
\lbrack n],a\in A}$ defined by three sets of conditions:\smallskip

\noindent in each row $i$ the entries are ordered the same way as in every $%
u_i^s$

\noindent $\sum_{i=1}^{n}x_{i}^{a}=0$ in each column $a$

\noindent $x_{1a}=0,x_{1b}=1$ where $a$ and $b$ are the outcomes ranked $k$
and $k+1$ by each $u_{1}^{s}\smallskip $

Note that $Q$ is non empty because it contains each matrix $U^{s}$.

By construction each $X$ in $Q$ defines a profile in ${\large \Pi }$ and $%
\lambda \cdot x_{i}^{\ast }=\lambda ^{i}\cdot x_{i}$ for all $i$. Therefore
we have%
\begin{equation*}
\sum_{i=1}^{n}\lambda ^{i}\cdot x_{i}<0\text{ for all }X\in Q
\end{equation*}%
\begin{equation*}
\lim_{s\rightarrow \infty }\sum_{i=1}^{n}\lambda ^{i}\cdot u_{i}^{s}=0\text{
for the sequence }U^{s}\text{ in }Q
\end{equation*}%
This is impossible: if the closed polyhedron $Q$ is disjoint from the
hyperplane $H:\sum_{i=1}^{n}\lambda ^{i}\cdot x_{i}=0$, it cannot contain
points arbitrarily close to $H$.

Thus there is some positive $\varepsilon $ s.t. for any profile $U$ in $%
{\large \Pi }$ we have $\sum_{i=1}^{n}\lambda \cdot u_{i}^{\ast
}<-\varepsilon $, and we can now conclude the proof in Case 1. These
inequalities imply for any profile $U$:%
\begin{equation}
\{\sum_{i=1}^{n}u_{i}=0\text{ and }u_{1k}^{\ast}<u_{1,k+1}^{\ast}\}%
\Longrightarrow \sum_{i=1}^{n}\lambda \cdot u_{i}^{\ast }\leq -\varepsilon
(u_{1,k+1}^{\ast}-u_{1k}^{\ast})  \label{21}
\end{equation}%
Indeed if $u_{1,k+1}^{\ast}-u_{1k}^{\ast}=1$ the profile $%
(u_{1}-u_{1k}^{\ast}\boldsymbol{1},u_{2}+u_{1k}^{\ast}\boldsymbol{1}%
,u_{3},\cdots ,u_{n})$ is in ${\large \Pi }$, and rescaling our profile by $%
\frac{1}{u_{1,k+1}^{\ast}-u_{1k}^{\ast}}$ implies the claim.

Note that in (\ref{21}) we can replace coordinate $1$ by any coordinate $i$.
Therefore we have%
\begin{equation}
\sum_{i=1}^{n}u_{i}=0\Longrightarrow \sum_{i=1}^{n}\lambda \cdot u_{i}^{\ast
}\leq -\frac{\varepsilon}{n} \sum_{i=1}^{n}(u_{i,k+1}^{\ast}-u_{ik}^{\ast})
\label{new}
\end{equation}

Because $\lambda _{k}>0$, the lottery $\mu $ obtained from $\lambda $ by
shifting $\frac{\varepsilon}{n}$ or $\lambda _{k}$, whichever is less, from $%
\lambda _{k}$ to $\lambda _{k+1}$ dominates $\lambda $, and property (\ref%
{new}) implies it is feasible.\smallskip

\textbf{Case 2 }$\lambda _{k}=0$

In this case, because $\widehat{z}\in \overleftrightarrow{W}$ it is a sum of 
$m$ elements $z^{j}\in W,j\in \lbrack m]$, each $z^{j}$ defined by $n$
utilities $(\overline{u}_{i}^{j})_{i=1}^{n}$ as in (\ref{22}). Note that if $%
k>1$ then $\overline{u}_{i_{0},k-1}^{j_{0}\ast }<\overline{u}%
_{i_{0},k}^{j_{0}\ast }$ for some $i_{0}, j_0$. Pick such $i_{0}$ and $j_{0}$
(or arbitrary ones if $k=1$). Let $a\in A$ be s. t. $\overline{u}%
_{i_{0},k}^{j_{0}\ast }=\overline{u}_{i_{0},a}^{j_{0}}$. For some small $%
\varepsilon >0$, modify $(\overline{u}_i^{j_0})_{i=1}^n$ to $%
(u_{i})_{i=1}^{n}$ by letting $u_{i_{0},a}=\overline{u}_{i_{0},a}^{j_{0}}-%
\varepsilon $, $u_{i_{1},a}=\overline{u}_{i_{1},a}^{j_{0}}+\varepsilon $ for
some $i_{1}\neq i_{0}$, and leaving all other utilities unchanged. Because $%
\lambda _{k}=0$ and by our choice of $i_{0}$,$j_{0}$, for small enough $%
\varepsilon $ we have $\sum_{i=1}^{n}\lambda \cdot u_{i}^{\ast }\geq
\sum_{i=1}^{n}\lambda \cdot \overline{u}_{i}^{j_{0}\ast }=\lambda \cdot
z^{j_0}=0$. As $\lambda $ is feasible, this must be an equality, and
therefore $z=\sum_{i=1}^{n}u_{i}^{\ast }\in Z$ and satisfies $z_{k}<z_{k+1}$
by construction. $\blacksquare $

\subsection{Proof of Lemma 4}

\noindent \textit{Statement If. }Pick\ two guarantees $\lambda ,\mu $ in $%
\mathcal{G}(n,p)$, such that $\lambda $ meets the property above while $\mu
\vdash \lambda $. Pick $k\in \lbrack p-1]$\textit{\ }and a profile $\pi $ as
in the statement. Choose a lottery ${\large \ell }$ implementing $\mu $ at $%
\pi $ and an agent $i$ reaching the maximum in (\ref{13}): we have $[{\large %
\ell }^{\ast i}]_{1}^{k}\leq \lbrack \mu ]_{1}^{k}\leq \lbrack \lambda
]_{1}^{k}$ and $[{\large \ell }^{\ast i}]_{1}^{k}=[\lambda ]_{1}^{k}$. As $k$
was arbitrary in $[p-1]$ we conclude $\mu =\lambda $ therefore $\lambda $ is
maximal.\smallskip

\noindent \textit{Statement Only If.} Suppose now that $\lambda \in \mathcal{%
G}(n,p)$ fails the property in the Lemma: there is some $k$ and some $%
\varepsilon >0$ s. t. at any profile $\pi $ there is some lottery ${\large %
\ell }$ implementing $\lambda $ at $\pi $ and such that%
\begin{equation}
\max_{i\in \lbrack n]}[{\large \ell }^{\ast i}]_{1}^{k}=[\lambda
]_{1}^{k}-\varepsilon  \label{14}
\end{equation}

We must show that $\lambda $ is not maximal. Suppose first $\lambda _{k}>0$
and construct $\lambda ^{\prime }$ dominating $\lambda $ by shifting a
weight $\delta $, smaller than $\varepsilon $ and $\lambda _{k}$, from $%
\lambda _{k}$ to $\lambda _{k+1}$ (and no other change). The lottery $%
\lambda ^{\prime }$ is still in $\mathcal{G}(n,p)$: at a profile $\pi $ the
lottery $\ell $ implementing $\lambda $ and meeting (\ref{14}) implements $%
\lambda ^{\prime }$ as well. Suppose next $\lambda _{k}=0$. Then we have for
all $i$

\begin{equation*}
\lbrack {\large \ell }^{\ast i}]_{1}^{k-1}\leq \lbrack {\large \ell }^{\ast
i}]_{1}^{k}\leq \lbrack \lambda ]_{1}^{k}-\varepsilon =[\lambda
]_{1}^{k-1}-\varepsilon
\end{equation*}%
so that if $\lambda _{k-1}$ is positive we can apply the previous argument.
If $\lambda _{k-1}=0$ again, we repeat this observation until we find some
positive $\lambda _{t}$, $t\leq k-2$, whose existence is assured by (\ref{14}%
). $\blacksquare $

\subsection{Proof of Theorem 1}

\noindent \textit{Step 1}. Recall the following notion from the
Shapley-Bondareva theorem. A family $S_{1},\ldots ,S_{m}$ of subsets of $[p]$
is \emph{balanced} if there exist positive weights $\gamma _{1},\ldots
,\gamma _{m}$ such that $\sum_{i:j\in S_{i}}\gamma _{i}=1$ for every $j\in
\lbrack p] $.\smallskip

\textbf{Lemma 5} Assume that\textit{\ }$p\leq 2n-2$\textit{, or }$p=2n$%
\textit{\ but }$n\neq 4,5$\textit{\ and let }$2\leq k\leq \lfloor \frac{p}{2}%
\rfloor $\textit{. Then there exists a balanced family }$S_{1},\ldots ,S_{m}$%
\textit{\ of subsets of }$[p]$\textit{\ of size }$k$\textit{\ each, such
that }$m\leq n$\textit{.}

Assume first that $p\leq 2n-2$ and $2\leq k\leq \lfloor \frac{p}{2}\rfloor $%
\textit{. }If $k$ divides $p$ the lemma is obvious (take a partition of $[p]$%
). Suppose $p=tk+r$ where $1\leq r\leq k-1$. Let $S_{i}=\{(i-1)k+1,\ldots
,ik\}$ for $i=1,\ldots ,t$. Also, let $S_{i}=C_{i}\cup \{tk+1,\ldots ,p\}$
for $i=t+1,\ldots ,t+k$, where the sets $C_{i}$ are of size $k-r$ and form
the $k$ cyclic intervals in a cyclic arrangement of $S_{t}$. Let $\gamma
_{1}=\cdots =\gamma _{t-1}=1,\gamma _{t}=\frac{r}{k},\gamma _{t+1}=\cdots
=\gamma _{t+k}=\frac{1}{k}$. These weights make $S_{1},\ldots ,S_{t+k}$ a
balanced family, and it remains to check that $t+k\leq n$.

We have $t+k<\frac{p}{k}+k\leq \max \{\frac{p}{x}+x:\,x\in \lbrack 2,\frac{p%
}{2}]\}=\frac{p+4}{2}$. If $p\leq 2n-2$ this gives $t+k<n+1$ as desired.

Assume next $p=2n$ and $2\leq k\leq n$. When $k$ divides $p$ a partition
works, so we may assume that $3\leq k\leq n-1$ and thus $n\geq 4$. We
further exclude the exceptional cases $n=4,5$ and assume $n\geq 6$. If $%
k\leq n-2$ we still have $\frac{p}{k}+k\leq n+1$ as in the original proof.
Thus we may assume that $k=n-1$. We provide two variants of the construction
of the balanced family, depending on parity.

\textit{Case 1. }$k=n-1$\textit{\ is even. }Partition $[p]=[2k+2]$ into $%
S,P_{1},\ldots ,P_{\frac{k}{2}+1}$ where $|S|=k$ and the other sets are
pairs. Take $S$ with weight $1$, and for each $P_{i}$, the union of all $%
P_{j}$, $j\neq i$, with weight $\frac{2}{k}$. This gives a balanced family
of size $\frac{k}{2}+2<n$.

\textit{Case 2. }$k=n-1$\textit{\ is odd. }Partition $[p]=[2k+2]$ into $%
S,T,P_{1},\ldots ,P_{\frac{k-1}{2}}$ where $|S|=k$, $|T|=3$ and the other
sets are pairs. Take $S$ with weight $1$, for each $P_{i}$ take the union of 
$T$ and all $P_{j}$, $j\neq i$, with weight $\frac{2}{k}$, and for each
element $a$ of $T$ take the union of $\{a\}$ and all the $P_{i}$ with weight 
$\frac{1}{k}$. This gives a balanced family of size $\frac{k-1}{2}+4\leq n$. 
$\blacksquare \smallskip $

\noindent \textit{Step 2}. Assume $(n,p)$ are as in Lemma 5 and let $2\leq
k\leq p-2$. Then for any $\lambda \in \mathcal{G}(n,p)$ we have $[\lambda
]_{1}^{k}\geq \frac{k}{p}$.

By duality, it suffices to show this for $2\leq k\leq \lfloor \frac{p}{2}%
\rfloor $. Let $S_{1},\ldots ,S_{m}$ with weights $\gamma _{1},\ldots
,\gamma _{m}$ be a balanced family as in the lemma. Consider a profile of
preferences in which $\{a_{j}:\,j\in S_{i}\}$ is the $k$-tail of the
preferences of agent $i$, $i=1,\ldots ,m$. Let $\ell $ be a lottery that
implements $\lambda $ at this profile. Then $1=\sum_{a\in A}\ell_
a=\sum_{i=1}^{m}\gamma _{i}\sum_{j\in S_{i}}\ell_{a_{j}}\leq
\sum_{i=1}^{m}\gamma _{i}[\lambda ]_{1}^{k}=\frac{p}{k}[\lambda ]_{1}^{k}$,
implying the desired inequality.\smallskip

\noindent \textit{Step 3. }We know from Proposition 3 and the maximality of $%
\lambda ^{vt},\lambda ^{rd}$ that $\mathcal{M}(n,p)$ contains the union of
the two intervals in the statement. Conversely we fix $\lambda \in \mathcal{G%
}(n,p)$, where $(n,p)$ are as in Lemma 5, and show that it is dominated by a
guarantee in those two intervals. We distinguish three cases.

\textit{Case 1. }$\lambda _{p}\geq \frac{1}{p}$. Set $\lambda _{p}=x$ and
keep in mind that feasibility implies $x\leq \frac{1}{n}$. We will show that 
$\lambda $ is dominated (weakly) by the guarantee $\mu \in \lbrack \lambda
^{uni},\lambda ^{rd}]$ s. t. $\mu _{p}=x$ : that is $\mu _{k}=x$ for $1\leq
k\leq n-1$ and $\mu _{k}=y$ for $n\leq k\leq p-1$, with $nx+(p-n)y=1$.

Set $p=n+q$ and partition $A$ as $\{a_{1},\cdots ,a_{n}\}\cup \{b_{1},\cdots
,b_{q}\}$ then consider a profile of preferences where for everyone:

the $a$-s occupy the ranks $1$ to $n-1$ and $p$ and each $a$ appears exactly
once in rank $p$;

the $b$-s occupy the ranks $n$ to $p-1$ and the pattern of the $b$-s is
cyclical for the first $q$ agents.

Pick a lottery ${\large \ell }$ implementing $\lambda $ at this profile.
Then ${\large \ell }_{a}\geq x$ for each $a$ implying $[\lambda
]_{1}^{k}\geq kx$ for $1\leq k\leq n-1$; moreover $\lambda _{p}=x$ by
assumption. It remains to show that $[\lambda ]_{p-r}^{p}\leq x+ry$ for $%
1\leq r\leq q-1$. Indeed by summing the implementation constraints for the
top $r+1$ outcomes of the first $q$ agents, we get (denoting the top outcome
of agent $i$ by $a_{i}$):%
\begin{equation*}
q[\lambda ]_{p-r}^{p}\leq \sum_{i=1}^{q}{\large \ell }_{a_{i}}+r%
\sum_{i=1}^{q}{\large \ell }_{b_{i}}=(\sum_{i=1}^{q}{\large \ell }%
_{a_{i}}+\sum_{i=1}^{q}{\large \ell }_{b_{i}})+(r-1)\sum_{i=1}^{q}{\large %
\ell }_{b_{i}}
\end{equation*}%
\begin{equation*}
\leq (1-(n-q)x)+(r-1)(1-nx)=q(x+ry)
\end{equation*}

\textit{Case 2. }$\lambda _{1}\leq \frac{1}{p}$. Set $\lambda _{1}=x$ and $%
p=n+q$. We show similarly that $\lambda $ is dominated (weakly) by the
guarantee $\mu \in \lbrack \lambda ^{uni},\lambda ^{vt}]$ s. t. $\mu _{1}=x$
: that is $\mu _{k}=x$ for $p-n+2\leq k\leq p$ and $\mu _{k}=y$ for $2\leq
k\leq q+1$, with $nx+qy=1$.

We consider a profile of preferences over the outcomes in $\{a_{1},\cdots
,a_{n}\}\cup \{b_{1},\cdots ,b_{q}\}$ where:

the $a$-s occupy the ranks $1$ and $p-n+2$ to $p$ and each $a$ appears
exactly once in rank $1$;

the $b$-s occupy the ranks $2$ to $q+1$ and the pattern of the $b$-s is
cyclical for the first $q$ agents.

Then the proof mimicks that in case 1 by showing first that a lottery
implementing $\lambda $ at this profile has $[\lambda ]_{p-k+1}^{p}\leq kx$
for $1\leq k\leq n-1$, then focusing attention on the first $q+1$ ranks to
show $[\lambda ]_{1}^{r+1}\geq x+ry$ for $1\leq r\leq q-1$. We omit the
details.\smallskip

\textit{Case 3. }$\lambda _{p}<\frac{1}{p}<\lambda _{1}$. Combining these
inequalities with those in step 2 we see that $\lambda $ is strictly
dominated by $\lambda ^{uni}$. $\blacksquare $

\subsection{Proof of Theorem 2}

\noindent We fix $1 \le q \le n$ such that $p = dn + q$ and prove the
statement by induction on $d$. It is clear for $d=1$ as $\{VT\}$ and $\{RD\}$
are the only two sequences and the intervals $[UNI(p),VT(n,p)]$, $%
[UNI(p),RD(n,p)]$ are in $\mathcal{M}(n,p) $.

Fix $d\geq 2$ and consider a sequence $\Gamma \in \{VT,RD\}^{d}$ starting
with $\Gamma ^{1}=VT$. By its definition (\ref{30}) the composition by $VT$
commutes with convex combinations of $\Gamma ^{2},\Gamma ^{2}\otimes \Gamma
^{3},\cdots $. Using the notation $VEX{\large [}\cdot {\large ]}$ for such
combinations we have 
\begin{equation}
VEX{\large [}VT,VT\otimes \Gamma ^{2},\cdots ,VT\otimes \Gamma ^{2}\otimes
\cdots \otimes \Gamma ^{d}{\large ]}=  \label{34}
\end{equation}%
\begin{equation*}
=VT\otimes VEX{\large [}UNI,\Gamma ^{2},\Gamma ^{2}\otimes \Gamma
^{3},\cdots ,\Gamma ^{2}\otimes \cdots \otimes \Gamma ^{d}{\large ]}
\end{equation*}%
where by the inductive assumption the second convex combination of canonical
guarantees in $\mathcal{C}(n,p-n)$ and of $UNI(p-n)$ is a maximal guarantee.
By Lemma 3 so is the left-hand convex combination $\lambda $, and by
Proposition 3 so is a convex combination of $UNI(p)$ and $\lambda $.

The proof of the inductive step for a sequence starting from $RD$ is more
involded, because $RD$ does not commute with convex combinations, even of
boundary lotteries: therefore property (\ref{34}) where $RD$ replaces $VT$
can only be true if the two sides use different convex combinations.

Observe first that if the boundary lottery $\lambda $ is maximal, $\lambda
\in \mathcal{M}(n,p-n)\cap \partial \Delta (p-n)$, then any $\mu =VEX{\large %
[}RD(n,p){\large ,}RD\otimes \lambda {\large ]}$ is in $\mathcal{M}(n,p)\cap
\partial \Delta (p)$ as well. That $\mu $ is on the boundary is clear. By (%
\ref{31}) $\mu $ takes the form 
\begin{equation*}
\mu =(\overset{n-1}{\overbrace{\frac{\alpha }{n},\cdots ,\frac{\alpha }{n}}}%
,(1-\alpha )\lambda ,\frac{\alpha }{n})
\end{equation*}%
Consider a profile similar to (\ref{33}) in the proof of Lemma 3, where by
maximality of $\lambda $ we choose $\pi $ ensuring property (\ref{13}) in
Lemma 4:%
\begin{equation*}
\begin{array}{cccccc}
\prec _{1} & a_{1} & \cdots & a_{n-1} & \overset{p-n}{\overbrace{\pi }} & 
a_{n} \\ 
\cdots & \cdots &  & \cdots & \pi & \cdots \\ 
\prec _{n} & a_{n} & \cdots & a_{n-2} & \pi & a_{n-1}%
\end{array}%
\end{equation*}%
If the lottery $\ell $ implements $\mu $ at this profile we have $\ell
_{a_{i}}=\frac{\alpha }{n}$ therefore its weight on the remaining $p-n$
outcomes in $\pi $ is $(1-\alpha )$ and the claim follows by Lemma 4 again.

We fix now an arbitrary convex combination 
\begin{equation*}
\Lambda =\sum_{j=2}^{d}\alpha _{j}RD\otimes \Gamma ^{2}\otimes \cdots
\otimes \Gamma ^{j}
\end{equation*}
in $\mathcal{G}(n,p)$ and claim that it takes the form $RD\otimes \lambda $
where $\lambda $ is some other convex combination 
\begin{equation*}
\lambda =\sum_{j=2}^{d}\beta _{j}\Gamma ^{2}\otimes \cdots \otimes \Gamma
^{j}.
\end{equation*}
This claim allows us to complete the induction step as follows. By the
induction hypothesis, $\lambda$ is in $\mathcal{M}(n,p-n)$, and it is easy
to see (and explained in detail below) that it is on the boundary. By what
we just observed, any $VEX{\large [}RD(n,p){\large ,}RD\otimes \lambda 
{\large ]}$ is also maximal; by the claim this means that any convex
combination of the guarantees corresponding to the initial subsequences of $%
\Gamma$ starting with $RD$ is maximal. Finally, Proposition 3 handles the
addition of the uniform guarantee.

\textit{Proof of the claim.} Recall that canonical guarantees are uniform on
their support, which we now describe for the canonical guarantees in our
sequence. We partition the ranks $1,\cdots ,p$ into subsets $S^{1},\cdots
,S^{d+1}$ each of size $n$ except for the last one of size $q$. The set $%
S^{1}$ is the support of $RD(n,p)$ (the ranks $1$ to $n-1$ and $p$). If $%
\Gamma ^{2}=RD$ then $S^{2}$ has the ranks $n$ to $2n-2$ and $p-1$, and the
support of $RD\otimes \Gamma ^{2}$ is $S^{1}\cup S^{2}$. If $\Gamma ^{2}=VT$
then $S^{2}$ has the rank $n$ and those from $p-n+1$ to $p-1$, and the
support of $RD\otimes \Gamma ^{2}$ is $S^{1}\cup S^{3}\cup \cdots \cup
S^{d+1}$ (the complement of $S^{2}$). Continuing in this fashion, each $%
\Gamma ^{j}$ defines a new set $S^{j}$ that is added to its support if $%
\Gamma ^{j}=RD$, while if $\Gamma ^{j}=VT$ we add $S^{j+1}\cup \cdots \cup
S^{d+1}$ to the support. We keep track of this construction by entering a
one for sets in the support and a zero for those outside it: with the
notation $\varepsilon \in \{0,1\}$ and $\varepsilon ^{\prime }=1-\varepsilon 
$ our sequence in $\mathcal{C}(n,p)$ is described by a table as follows%
\begin{equation*}
\begin{array}{cccccccc}
& S^{1} & S^{2} & S^{3} & S^{4} & \cdots & S^{d} & S^{d+1} \\ 
RD\otimes \Gamma ^{2} & 1 & \varepsilon _{2} & \varepsilon _{2}^{\prime } & 
\varepsilon _{2}^{\prime } & \cdots & \varepsilon _{2}^{\prime } & 
\varepsilon _{2}^{\prime } \\ 
RD\otimes \Gamma ^{2}\otimes \Gamma ^{3} & 1 & \varepsilon _{2} & 
\varepsilon _{3} & \varepsilon _{3}^{\prime } & \cdots & \varepsilon
_{3}^{\prime } & \varepsilon _{3}^{\prime } \\ 
\cdots & \cdots & \cdots & \cdots & \cdots & \cdots & \cdots & \cdots \\ 
RD\otimes \Gamma ^{2}\otimes \cdots \otimes \Gamma ^{d} & 1 & \varepsilon
_{2} & \varepsilon _{3} & \varepsilon _{4} & \cdots & \varepsilon _{d} & 
\varepsilon _{d}^{\prime }%
\end{array}%
\end{equation*}%
where $\varepsilon _{j}=1$ if $\Gamma ^{j}=RD$, $\varepsilon _{j}=0$ if $%
\Gamma ^{j}=VT$.

Defining $\Theta _{k}=n\sum_{j=2}^{k}\varepsilon _{j}+(p-kn)\varepsilon
_{k}^{\prime }$ we see in the table that $\Theta _{k}$ is the size of the
support of $\Gamma ^{2}\otimes \cdots \otimes \Gamma ^{k}$, while that of $%
RD\otimes \Gamma ^{2}\otimes \cdots \otimes \Gamma ^{k}$ has cardinality $%
\Theta _{k}+n$. On its support $RD\otimes \Gamma ^{2}\otimes \cdots \otimes
\Gamma ^{k}$ is worth $\frac{1}{\Theta _{k}+n}$ while $\Gamma ^{2}\otimes
\cdots \otimes \Gamma ^{k}$ is $\frac{1}{\Theta _{k}}$ on its own support.

Clearly, but critically, there is a column with only zeroes: this holds if $%
\varepsilon _{2}=0$ ($\Gamma ^{2}=VT$), or if $\varepsilon _{2}=1$ but $%
\varepsilon _{3}=0$, etc.., until, if $\varepsilon _{j}=1$ for all $j$, the
last column is null. A symmetric argument shows that in addition to the
first column, there is another column full of ones. The first remark implies
that $\Lambda $ and $\lambda $ are respectively in $\partial \Delta (p)$ and 
$\partial \Delta (p-n)$; the second that the maximal coordinate of $\lambda $
is $\lambda _{+}=\sum_{j=2}^{d}\frac{\beta _{j}}{\Theta _{j}}$. Now we
select the coefficients $\beta _{j}$ such that%
\begin{equation*}
\frac{1}{n\lambda _{+}+1}\frac{\beta _{j}}{\Theta _{j}}=\frac{\alpha _{j}}{%
\Theta _{j}+n}\text{ for all }j=2,\cdots ,d\text{, and }\sum_{j=2}^{d}\beta
_{j}=1
\end{equation*}%
Check that $\beta $ is well defined because summing the first $d-1$
equalities above implies%
\begin{equation*}
\frac{n\lambda _{+}}{n\lambda _{+}+1}=\sum_{j=2}^{d}\frac{n}{\Theta _{j}+n}%
\alpha _{j}<1
\end{equation*}%
which determines $\lambda _{+}$. After rearranging the equation above as%
\begin{equation*}
\frac{1}{n\lambda _{+}+1}=\sum_{j=2}^{d}\frac{\Theta _{j}}{\Theta _{j}+n}%
\alpha _{j}
\end{equation*}%
the last equality $\sum_{j=2}^{d}\beta _{j}=1$ follows.

We check finally the equality $\Lambda =RD\otimes \lambda $ for this choice
of $\beta $. Because $\lambda \in \partial \Delta (p-n)$ the lottery $%
RD\otimes \lambda $ is given by (\ref{31}): in particular it is constant on
each set $S^{k}$, just like $\Lambda $. We see in the table that $RD\otimes
\lambda $ equals $\frac{\lambda _{+}}{n\lambda _{+}+1}$ in $S^{1}$, while $%
\Lambda $ is worth $\sum_{j=2}^{d}\frac{\alpha _{j}}{\Theta _{j}+n}$ so they
coincide. Each entry in another column $S^{k}$ at row $j$ adds $\varepsilon 
\frac{1}{n\lambda _{+}+1}\frac{\beta _{j}}{\Theta _{j}}$ to $RD\otimes
\lambda $ and $\varepsilon \frac{\alpha _{j}}{\Theta _{j}+n}$ to $\Lambda $,
where $\varepsilon $ is the coefficient of that particular entry, so the
desired equality follows. $\blacksquare $

\end{document}